\documentclass[manuscript,screen]{acmart} 
\AtBeginDocument{%
  \providecommand\BibTeX{{%
    \normalfont B\kern-0.5em{\scshape i\kern-0.25em b}\kern-0.8em\TeX}}}
\usepackage{array, booktabs, xltabular}
\usepackage{color,soul}
\newcommand{\new}[1]{\textcolor{black}{#1}}

\setcopyright{acmcopyright}
\copyrightyear{2024}
\acmYear{2024}
\acmDOI{XXXXXXX.XXXXXXX}

\acmConference[Under submission]{}{2024}{}
\acmBooktitle{Under submission.}
\acmPrice{15.00}
\acmISBN{978-1-4503-XXXX-X/18/06}

\acmSubmissionID{123-A56-BU3}

\begin{document}

\title[Human-AI collaboration is not very collaborative yet]{Human-AI collaboration is not very collaborative yet: A taxonomy of interaction patterns in AI-assisted decision making from a systematic review}

\author{Catalina Gomez}
\affiliation{\institution{Johns Hopkins University} \country{USA}}
\author{Sue Min Cho}
\affiliation{\institution{Johns Hopkins University} \country{USA}}
\author{Shichang Ke}
\affiliation{\institution{Johns Hopkins University} \country{USA}}
\author{Chien-Ming Huang}
\affiliation{\institution{Johns Hopkins University} \country{USA}}
\author{Mathias Unberath}
\affiliation{\institution{Johns Hopkins University} \country{USA}}


\renewcommand{\shortauthors}{C. Gomez et al.}

\begin{abstract}
Leveraging Artificial Intelligence (AI) in decision support systems has disproportionately focused on technological advancements, often overlooking the alignment between algorithmic outputs and human expectations. A human-centered perspective attempts to alleviate this concern by designing AI solutions for seamless integration with existing processes. Determining what information AI should provide to aid humans is vital, a concept underscored by explainable AI's efforts to justify AI predictions. However, how the information is presented, e.g., the sequence of recommendations and solicitation of interpretations, is equally crucial as complex interactions may emerge between humans and AI. While empirical studies have evaluated human-AI dynamics across domains, a common vocabulary for human-AI interaction protocols is lacking. To promote more deliberate consideration of interaction designs, we introduce a taxonomy of interaction patterns that delineate various modes of human-AI interactivity. We summarize the results of a systematic review of AI-assisted decision making literature and identify trends and opportunities in existing interactions across application domains from 105 articles. We find that current interactions are dominated by simplistic collaboration paradigms, leading to little support for truly interactive functionality. Our taxonomy offers a tool to understand interactivity with AI in decision-making and foster interaction designs for achieving clear communication, trustworthiness, and collaboration.

\end{abstract}




\received{20 February 2007}
\received[revised]{12 March 2009}
\received[accepted]{5 June 2009}

\maketitle

\section{Introduction}
Advances in Artificial Intelligence (AI) developments open new possibilities for supporting human decision making across a wide variety of applications. 
\new{Decision making tasks in a broad range of applications share a process that starts when evidence is presented before making a decision within discrete choices, usually with follow-up effects. 
Within this framework, the decision-making process emerges as a scenario for human-AI teamwork where at a minimum two parties, i.e., the human and the AI, factor into finding a solution to the decision problem. The exact dynamics of how this collaboration occurs can vary from one situation to another, leading to multiple interaction options that range from simple recommendations to involved exchanges~\cite{bansal2019beyond,lai2023towards,bertrand2023selective}. 
To bridge algorithmic suggestions and human expectations, embracing a human-centered approach in designing AI solutions is crucial to identify \textit{what} information AI should provide to aid humans while ensuring a safe and transparent operation. 
It is equally crucial to understand \emph{how} and \emph{when} to best communicate the information for designing successful human-AI interactions. 
Explainable AI (XAI) expands model capabilities by providing not just so-called black box recommendations but also justifications tailored to end users' needs. The design of justifications from a human-centered perspective informs what AI models should deliver and it naturally involves careful thought on how to present this support, although it may not specifically address the ensuing interactions.} 
The presentation style (\textit{how}) and the strategic timing (\textit{when}) for providing AI-generated insights are closely related to the type and sequence of interactions between humans with AI, which are ultimately enabled by the affordances of certain assistance elements~\cite{lai2023towards}. 
Despite the importance of and opportunities in interaction design within AI, there is not currently a common vocabulary to describe and differentiate these interactions.

Interactivity is a familiar concept to humans and widely studied in more specialized domains such as information visualization, interface and software design~\cite{yi2007toward}, and human-human interactions from a social perspective~\cite{magnusson_2018}. 
Understanding interactions requires delving into multiple dimensions that involve subjects, modes, and purposes of interaction, and the context in which they take place~\cite{schleidgen2023concept}.
Among these, interaction patterns emerge as sequences of behaviors that occur more often than by chance between agents and systems or artifacts.
We have learned from these disciplines the importance of deliberate choices in selecting interaction types crucial for achieving specific goals, rather than imbuing unnecessary high levels of interactivity that do not result in better products~\cite{sims1997interactivity}.
\new{Likewise, finding the right balance of interactivity between humans and AI systems is not just a matter of enhancing user experience but is essential for achieving clear communication, trustworthiness, and meaningful collaboration.}
Previous AI design guidelines consider pertinent aspects during interactions with end users, including the time when to act or interrupt, display information relevant to the user's current task, and deliver an experience aligned with expectations and avoid bias reinforcement~\cite{amershi2019guidelines}. As noted from these aspects, implementing interactivity requires an understanding of capabilities of both agents and materialization through an interface. 
\new{While current AI systems excel in offering problem-solving capabilities, there is often a disproportionate emphasis on the technological advancements, overlooking the critical aspects of user interface and experience. This oversight is apparent in many empirical studies where interactions with AI agents are typically reduced to basic actions like menu selections or button clicks.}

Which forms of interactivity to incorporate in human-AI interactions is an open question and may depend on the overall context, emphasizing the need to deliberately study interaction patterns between humans and AI that can guide the development of better solutions.
Researchers in numerous empirical studies have actively evaluated human-AI interactions across various application domains and decision tasks~\cite{lai2023towards,bertrand2023selective}. However, the specific configurations needed to evaluate the effect of different AI assistance elements (or other context-related factors) on humans interacting with AI, can lead to multiple forms of collaboration or the actual interactions afforded. 
Referring to different types of interactions is particularly difficult since we lack of a common language that captures a broad range of ways in which humans can interact with AI when solving decision making tasks. The absence of a comprehensive taxonomy hampers the aggregation and synthesis of knowledge across diverse studies and limits the exploration of new interaction paradigms. 
Meanwhile, the field of Human-Robot Interaction (HRI) has studied in particular how humans and robots communicate, collaborate, and engage with each other, often through the analysis of observable patterns in their interactions~\cite{david2022interaction,sauppe2014design,ma2019design}. 
These patterns cover various aspects of social interactions, such as greetings, attention, feedback, turn-taking, social cues, and farewell. All these elements foster a more natural and efficient communication experience, which can take place in AI technologies that incorporate natural language abilities. 
Other domains such as information visualization and user interface design have also developed taxonomies and libraries of interaction patterns based on different criteria such as user’s intent, purpose, scope, abstraction level, and granularity \cite{yi2007toward,silva2020classifying}.
These categorizations serve as a base to build on more comprehensive taxonomies grounded on and informed by evidence from empirical studies and incorporating knowledge from interaction design in other disciplines.  
In the context of AI, previous attempts to describe possible interactions between humans and machines or AI have grouped them by user control and initiative~\cite{van2021human,cheng2022mapping}, task nature~\cite{parasuraman2000model}, and level of automation~\cite{Mackeprang2019-dz}. Overall, the existing classifications of interactions that involve AI are mostly focused on specific domains or contexts but do not provide a general and comprehensive framework for characterizing and analyzing interaction patterns that apply to multiple domains. 
This highlights the importance of a structured framework for categorizing different interaction patterns, providing designers not only with an understanding of existing ways for humans to collaborate with AI but also to make informed decisions on interaction design to build better partnerships between humans and AI.
 
To address this gap, we first conducted an extensive systematic review on human-AI interactions that have been reported for human-AI decision making scenarios. We searched for relevant articles from five databases that cover Human-Computer Interaction (HCI) studies and related disciplines and selected 105 to conduct a detailed coding and analysis of the sequences of interactions that exist between humans and AI. Furthermore, we considered the task context, AI system involved, and the evaluation methods included in these empirical studies. 
Grounded in the trends from our systematic review, we propose a taxonomy of interaction patterns that comprises seven interaction patterns that arise between humans and AI. 
To the best of our knowledge, there is not yet a comprehensive and structured classification of existing interaction patterns between humans and AI. We propose our taxonomy of interactions as a tool to better understand existing interaction patterns across domains and applications, allowing us to identify the occurrences of common interactions across multiple domains. We envision the use of taxonomy to foster dialogues on interactivity in AI-assisted decision making, encourage refinement and evaluation of novel patterns, and ultimately design better and more user-centered AI-based solutions.


\section{Methods}
\subsection{Search strategy and selection criteria}
This survey focuses on Human-AI interaction paradigms for explicit decision-making tasks, in contrast to proxy task where users are asked to simulate the AI outputs. Therefore, we aim to understand and evaluate the works that study human-AI interactions during decision-making tasks under AI assistance, instead of improvements of the model. 
Our survey covers studies conducted between 2013 and June 2023. Specifically, we searched within five databases: ACM Digital Library, IEEE Explore, Compendex, Scopus, and PubMed. The first four have extensive coverage of relevant studies in HCI covering conference proceedings and journal publications (Compendex and Scopus included papers from more subjects), while PubMed allowed us to capture research specifically related to medical applications of Human-AI interaction and decision-making.
We defined the search terms covering four dimensions: use of AI systems, human-AI interaction or collaboration studies, decision-making tasks, and interaction design. We included the last term since we wanted to focus on articles that evaluate interactions with AI systems during decision making tasks. The complete set of keywords used in our search can be found in the appendix.
We defined the following inclusion criteria:
\begin{itemize}
    \item The tasks in consideration are those related to decision-making, and in particular, we limit the study selection to those that implement complete decision making processes and not only evaluations of decision makers' perceptions, such as understanding, preferences, or judgments of AI's advice.  
    \item The paper shows an implementation of the interface that was presented to human users to interact with AI. 
    \item The modes of interaction encompass screen-based interfaces, virtual agents, and non-embodied setups.
    \item We have restricted our selection solely to papers featuring empirical user studies. 
\end{itemize}

In addition to the inclusion criteria above, our search excluded studies in robotics and gaming by filtering out these keywords in the title and abstract. We excluded studies that involve robots because physical embodiment enables more dimensions of an interaction and those involving gaming scenarios because they are more complex, with less constraints to study how humans can interact with AI assistance. However, studies that implement decision making tasks through gamified tasks were included.
Other survey papers or comments were also excluded by filtering out keywords in the title and abstract. 

\subsection{Study selection}
The initial search returned 3,770 papers, and 358 duplicates were found and deleted automatically. This left us with a total of 3,412 papers to screen. They were assigned to two authors to first go through title and abstract screening, followed by full-text screening. The screening phase was oriented towards the exclusion of papers that were not focused on human-AI interaction, i.e., limited to technical contributions, did not involve a complete decision-making task, were short papers (less than 8 pages), involve gaming or robotics, have not been peer-reviewed, and were review or survey papers.
We did not constrain our selection to works that directly manipulate the type of interaction between the human and AI. Our main interest was on which were the existing/available ways for humans to interact with AI agents in the evaluation of human-AI decision making. 
The total and abstract screening excluded 2,893 papers, and at the full-text review stage, 363 papers were filtered out. Lastly, 156 were considered for the information extraction stage, of which 51 were removed as a more detailed reading allowed us to identify that they did not satisfy the inclusion criteria of supporting actual AI-assisted decision making tasks. Figure \ref{fig:prisma} in the Appendix summarizes the complete paper selection process. At the end, 105 articles were included in our review.

\subsection{Data extraction strategy}
\subsubsection{Analysis process}
The data extraction template was developed by all authors and informed by previous surveys of empirical studies in human-AI interaction~\cite{bertrand2023selective,lai2023towards}. 
Two authors distributed the final selected articles to be analyzed and coded the assigned articles independently. Then, one author checked the individual reports of each article to ensure consistency in the final extraction. Further discussions with the other authors took place to clarify discrepancies in the interactions or ambiguous cases. 
For the analysis of the interaction patterns, the authors reviewed the sequences of interactions and discussed how to group them into the design patterns that were repeated and are presented in this work. We iterated over the definition of each pattern to refine the actual components that constitute the interaction. 

\subsubsection{Context}
We identified general information in which the decision making task takes place. This includes the domain and we adapted the categories initially proposed in this survey of AI-assisted decision making~\cite{lai2023towards}. Furthermore, we specified the decision making task to be completed by the human (e.g., detection of hate speech, sleep stage classification, price estimation, among others) and the level of expertise required to successfully complete the task. 

\subsubsection{AI system}
As we are interested in humans interactions with AI agents, we retrieved the original goal of incorporating AI assistance in the decision making task, and briefly characterized the AI system used in the study. In particular, we extracted the technique supporting the AI's recommendations (whether a real model was used or the outcomes were simulated), its performance (if any evaluation metric was reported), and the terminology used to introduce the AI agent to participants in the user study. Further details such as data type and source, output type were not reported as the main focus of this survey is on the interactions rather than the type of AI methods as previously surveyed in~\cite{lai2023towards}.

\subsubsection{Evaluation}
In the empirical studies selected for this survey, humans were directly interacting with an AI systems and it is relevant to evaluate how they perceive such assistance and the outcomes of this interaction. We extracted the constructs evaluated in the studies and classified them as objective or subjective dependent variables depending on how they were assessed.

\subsubsection{Interaction building blocks}
An interaction involves a reciprocal action or influence between two agents in the context of this survey~\cite{schleidgen2023concept}.
To characterize this, we defined two elements: the action undertaken and the resulting output of that action. These two elements constitute the interaction building blocks that can be integrated into more complex interactions. In the definition of our taxonomy of human-AI interactions, we considered these building blocks as the main elements that constitute the interaction patterns. 
Drawing insights from prior studies on human-AI interactions, we defined the following (action - output) pairs available for the agents involved in the interaction. 
\new{These building blocks primarily focus on the AI assisting the user in decision-making processes and are grouped based on the main action and specify the agent that can execute it. It is important to highlight that these building blocks can also occur sequentially, where one may be triggered in response to another. However, we have chosen to maintain a granular level of detail in the following descriptions to better capture the nuances of individual actions that later compose the interaction patterns.}
\begin{itemize}
    \item Predict - Outcome: The agent produces a solution to the primary decision-making task after receiving task information. This action is observed to be executed by either the AI or the user independently. 

    \item Decide - Outcome: The agent integrates the assistance they received with the task-related information available to finalize the decision outcome. This action is typically observed to be executed by the user.

    \item Provide - Options: The agent offers solutions for a secondary task that, while not directly resolving the primary decision-making task, are still informative. This action is typically observed to be executed by the AI. 

     \item Display - Information: The agent presents supplementary evidence (e.g., explanations, uncertainty values, alternate solutions) supporting a solution to the primary decision-making task. This action is observed to be predominately executed by the AI.

    \item Request - Outcome/Information: The agent actively seeks information or solutions from its counterpart. This action is observed to be either mandatory or optional and is typically executed by the AI when it requires user inputs, or by the user when they seek a direct solution or supplementary information from the AI. 

    \item Collect - Inputs: The agent gathers task-related information and provides it to the other agent. This action is observed to be typically executed by the user when their input is needed for the AI to provide a solution to the decision-making task.

    \item Modify - Outcome/Information: The agent makes changes to the solutions or supplementary information provided by its counterpart. This action is typically observed to be optional and executed by the user. 

    \item Delegate - Decision: The agent decides whether to retain responsibility for the task or transfer to its counterpart. This action is observed to be executed by either the AI or the user. Events after delegation can differ, ranging from complete surrender of agency to opportunities for supervising the other agent's decision-making process. 
    \item Other: if an action does not fit the previous types.      
\end{itemize} 

These concepts were refined and iterated as we reviewed more works. For each paper, we first identified all possible (action - output) pairs as the building blocks for each agent involved in the decision-making task and included a brief description in free text form. Then, we defined a sequence of interactions considering the order in which these events take place and the agent in charge. The sequences were not preset in advance in the extraction template since we wanted to discover the interaction patterns here. We note that depending on the experimental manipulation of the user study, different modes of interactivity with the user could be plausible and we separated these into different sequences.

\section{Results}
\subsection{Taxonomy of interaction patterns for AI-assisted decision making}

\begin{figure}
    \centering
    \includegraphics[width=0.95\linewidth]{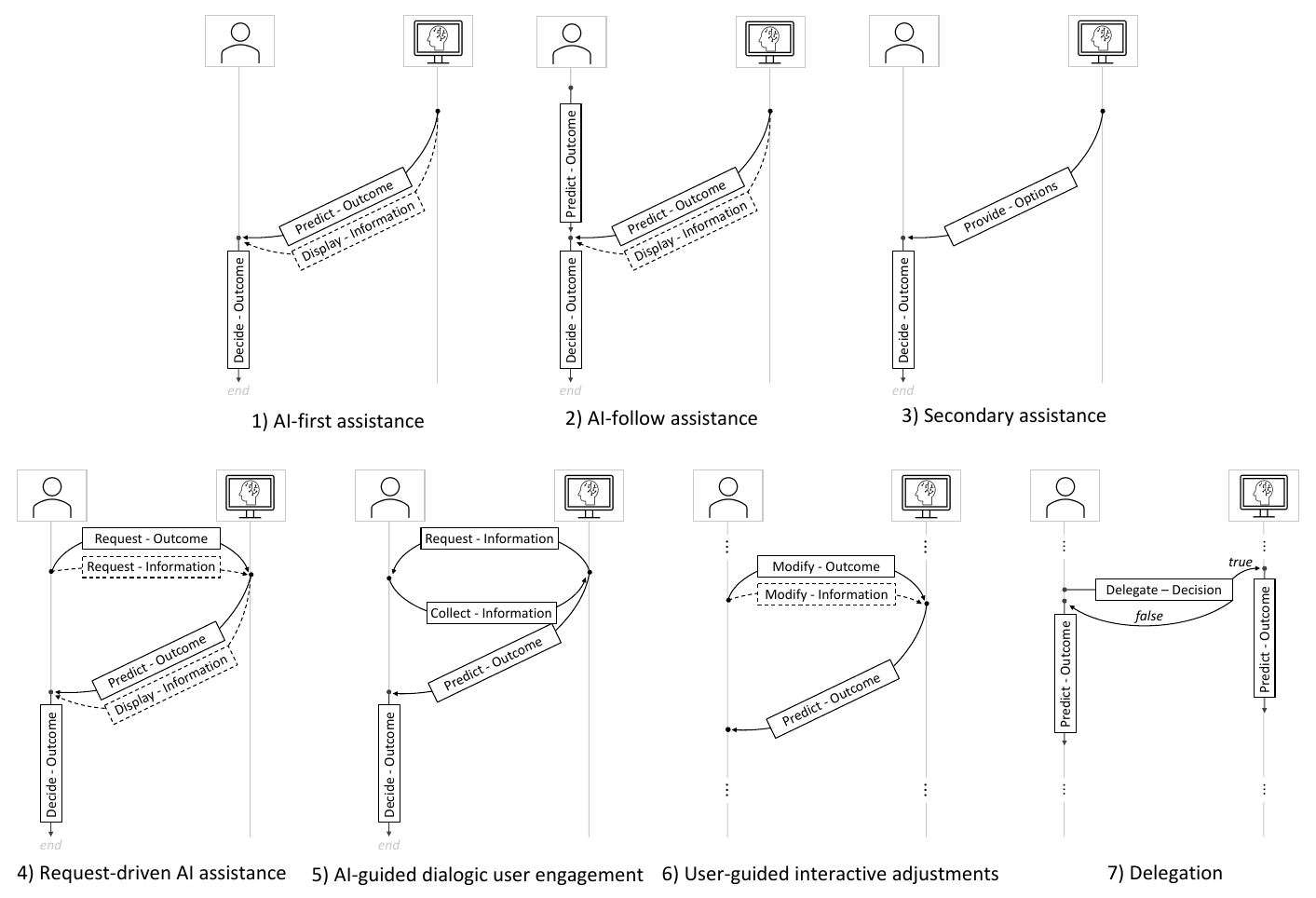}
    \caption{Taxonomy of interaction patterns identified in AI-assisted decision making. The user (human) and the AI are represented as separate agents and the temporal evolution of the interactions is illustrated from top to down. The boxes contain the building blocks (action-output) pairs that compose each pattern. The direction of arrows denote the agent who started the action. Dashed lines represent optional operations.}
    \label{fig:patterns}
\end{figure}

We present seven categories of interaction patterns that we have identified in our corpus, illustrated in Figure~\ref{fig:patterns}. \new{To formulate the taxonomy, we began by reviewing previous literature containing taxonomies of interaction techniques in other domains, such as information visualization~\cite{yi2007toward}, human-robot interaction~\cite{sauppe2014design}, multi-agent systems~\cite{cabri2002modeling}, and educational technologies~\cite{sims1997interactivity}}. Contrasting the concepts in these taxonomies with a sample of studies in human-AI interaction that we were familiar with, we identified potential ways in which the types of interactions described previously could potentially apply to humans interacting with AI during decision making. 
In each pattern formulation, we considered an appropriate level of abstraction so that they can capture multiple actions of the agents and generalize over various studies included in the systematic review. 
\new{The interaction patterns we defined in the taxonomy are comprised of the interaction building blocks and can borrow some of the other pattern categories. 
We identified the different categories of interaction patterns using a combination of an automated search for certain action-output pairs present in sequences of interactions that we characterized in the data extraction stage of the review, followed by manual inspection and verification.} 
Interactions involve changes over time and we attempted to capture this evolution/progression in the interaction patterns presented below and in the diagrams that illustrate them. 
Furthermore, we included a ``Other patterns'' category to present the interaction patterns that did not exactly fit into the main categories. 
In the following descriptions, we assume that users already have some background knowledge or intuitions that can be used during the decision making task.
Lastly, the classification of interaction patterns does not mean they are mutually exclusive events, but elements that can be consolidated and combined.  

\paragraph{AI-first assistance}
This pattern manifests when the decision-making problem and the AI-predicted outcome are simultaneously displayed to the user. As the ultimate decision maker, the user can choose to incorporate the AI's advice into their final decision or opt to disregard it. When task-related stimuli (e.g., images or case details) are presented alongside the AI-predicted outcome, the user is provided with a more comprehensive set of information to consider. \new{In addition, the AI's outcome can be accompanied by support information as captured in the Display - Information building block.}
This pattern has been previously observed and referred to as the "concurrent paradigm" ~\cite{tejeda2022ai} or "one-step workflow" ~\cite{Fogliato2022-uw}. 


\paragraph{AI-follow assistance}
This pattern begins with the user forming an independent preliminary prediction given the decision-making problem. Following this initial judgment, the AI's predicted outcome is presented and may accompanied by support information. This procedure provides the user with a reference (their initial assessment) to compare against the AI's advice, and an opportunity to reassess their initial judgment. This approach has been identified as the ``sequential paradigm'' ~\cite{tejeda2022ai} or ``two-step workflow'' ~\cite{Fogliato2022-uw} and has been commonly used to evaluate the human's reliance on the AI's advice.


\paragraph{Secondary assistance}
In this pattern, the AI offers information that does not serve as direct solutions to the decision-making problem. The user must interpret this supplementary information as an auxiliary task to determine its relevance and decide how to incorporate it into their primary decision-making process. We distinguish this as a unique interaction pattern because users respond differently to direct assistance compared to more secondary assistance in their decision-making process. For example, machine learning models can predict risk values associated with certain profile information and the human's decision problem is to make an investment decision~\cite{Dikmen2022-yj}.

\paragraph{Request-driven AI assistance}
In this interaction pattern, the user has to actively seek information or solutions from the AI. Rather than the AI's inferences being automatically presented, the user can control when they want to receive the AI assistance. Meanwhile, the user can spend more time deliberating about the problem, a strategy known as cognitive forcing~\cite{Bucinca2021-cv, Park2019-oo}. 
This pattern can be perceived as less intrusive to the user, as it empowers the user to ``ask'' for information or solutions from the AI, and allows the user to anticipate the AI's assistance in the decision-making process. 

\paragraph{AI-guided dialogic user engagement}
Within this interaction pattern, the AI facilitates a dialogue-like engagement with the user. Guided by the AI's instructions, the user responds by providing pertinent information. The iterative exchange continues until the AI's instruction requirements are satisfied, and is followed by the presentation of the AI's predicted outcome for the decision-making task. This responsive exchange not only involves retrieving and sharing information in line with the task but also ensures that the users recognize the influence of their inputs on the AI's predicted outcome. While this pattern has been commonly observed in humans interacting with conversational agents \cite{Jiang2022-kt, Gupta2022-bf}, it is not only limited to traditional conversational interfaces~\cite{Gomez2023-la}.

\paragraph{User-guided interactive adjustments}
Inspired by the taxonomy of interactive explanations recently proposed from a scoping review~\cite{bertrand2023selective}, we included an interaction pattern where humans can modify the outcome space of the AI agent. Typically, information flows from the AI to the user. However, in this pattern, the direction of flow is reversed, with humans providing the AI with feedback, corrections, or information to shape its inferences. 
While a detailed classification of potential modifications is beyond the scope of this survey, we distinguished cases in which the changes are merely visual updates in the interface or considered as feedback to improve the underlying AI models, as in interactive machine learning~\cite{amershi2014power}.

\paragraph{Delegation}
In this interaction pattern, both the user and the AI leverage their unique strengths and capabilities to optimize the decision-making outcome. Delegation can be a strategic choice when one agent assesses its counterpart as better equipped for a particular task ~\cite{Fugener2022-fh}. On the other hand, if an agent feels confident in their ability to complete the task, they will take the lead. Studies highlight that the complementary abilities of humans and AI when synergized properly, can enhance the decision-making outcome ~\cite{Zhang2022-tx}.

\paragraph{Others}
We included within this category those patterns that involve a combination of the main interaction blocks and did not fit into the patterns described before. More complex interaction emerges when the decision making problem may involve multiple individual decisions, agents (more than two), and continuous interactions with an AI agent.


\subsection{Identification of interaction patterns in AI-assisted decision making studies}
We identified the different categories of interaction patterns in the selected 105 articles. If a study included more than one interaction sequence, i.e., the experimental manipulation resulted in different ways in which users can interact with the AI, we considered them separately and counted patterns in each one. In total, we analyzed 131 sequences. 
The most common interaction pattern during empirical evaluations of AI-assisted decision making tasks was the AI-first assistance (n=67), followed by the AI-follow assistance (n=28). Furthermore, the AI's solutions to the decision making task were presented along with additional information in the majority of the cases, 81\% and 68\% during AI-first and AI-follow interactions, respectively.  
\new{For instance, in the AI-first pattern for predicting student outcomes, the AI takes student-related features and predicts pass or fail with a confidence value~\cite{Rastogi2022-dh}. Participants then review this AI prediction alongside the student data to make their final decision. Meanwhile, in the AI-follow pattern for another binary task, participants make a initial prediction regarding the output of speed dating events, and then a final one after seeing AI's prediction. }
We observed 16 instances of the Secondary assistance pattern in which the AI support did not propose a direct solution to the decision making task but rather estimated an outcome informative for the task. In particular, we noticed that most of the decision making tasks required expertise (11/16). 
\new{As an example, in the task of predicting gestational diabetes mellitus (GDM), healthcare professionals are presented with various descriptive features such as a history of diabetes mellitus, age, body weight, etc~\cite{Du2022-od}. The AI system processes this information and outputs a categorized risk of GDM, labeling it as either low or high. Rather than offering a direct solution, the AI presents this risk category alongside an explanation, thereby serving as a form of secondary assistance.}

Sometimes during the decision making task, users had to actively seek support from the AI agent as specified in the Request-driven AI assistance (n=25). More specifically, the requests could be for a direct a solution to the decision making task (n=14) or for the presentation of support information (n=13). In the former, only in three cases the request for the AI's solution was optional~\cite{Kumar2021-cm,Tolmeijer2021-dr,Baudel2021-ad}, meaning that users could come with a solution to the decision making task on their own. 
\new{As illustrated in a house search scenario, users could choose to use an AI system to help them find a house that satisfies certain requirements, with the option to directly submit the suggested house or verify~\cite{Tolmeijer2021-dr}.}
Meanwhile, the support information at the user's discretion was identified in eight cases~\cite{Calisto2022-bx,Vossing2022-si,Suresh2020-mx,van2021moral,Liu2021-gm,Molina2022-ms,Prabhudesai2023-he}, for demanding explanations in particular. 
In the AI-first assistance pattern, where the user may not have an opportunity to form an independent assessment of the decision making problem, we observe cases (n=10) in which users are given the ability to control when they want to receive the AI assistance via a request~\cite{Khadpe2020-vy,Mackeprang2019-dz,Baudel2021-ad,Molina2022-ms,Gomez2023-la,Bucinca2021-cv}.  

To a lesser extent, we found the interaction patterns that involve more exchange components between the human and the AI agent. For the AI-guided dialogic user engagement, as the name suggests, five out of the six interactions were supported via conversations with the AI agent. Through conversational interfaces, users had to provide some constraints given in the decision making problem for the AI to propose a candidate solution ~\cite{Khadpe2020-vy,Jiang2022-kt,Gupta2022-bf}. 
However, such exchange of information does not necessarily rely on a conversational interface, as demonstrated in the evaluation of an AI system that could constantly provide guidance on a decision sub-problem for identifying bird categories~\cite{Gomez2023-la}. \new{The two-way interaction occurs as the AI requests and suggests bird attributes for description, culminating in a bird category suggestion. Users can actively engage by processing the attributes, considering the AI's input, and making decisions regarding the bird's attributes. Ultimately, they verify the AI's suggested bird category.}
For a different purpose, interactivity components also support the adjustment of the AI's outcome space, and we found the User-guided interactive adjustments in nine cases that differed on the observed effect of the adjustment. For instance, manipulations of the inputs result in new AI's outcome computations for exploratory purposes~\cite{Liu2021-gm,Gu2023-zl,Nguyen2018-ao,Zytek2022-hr,Suresh2022-xp}. \new{Such functionality can be incorporated into interactive explanations, where users can manipulate input values of a specific instance and observe the change in recidivism predictions~\cite{Liu2021-gm}.}
When adjustments did not directly translate into an updated AI's outcome, users' feedback was considered for future improvement of the model~\cite{Ashktorab2021-us,Molina2022-ms,Smith-Renner2020-hs,Lee2021-lu}. \new{Manual labeling after observing AI predictions can be leveraged to identify incorrect intent classifications from textual samples and re-train the models~\cite{Ashktorab2021-us}.}

An opportunity to delegate decisions was observed in nine sequences of interactions, though with differences in the conditions for delegation. 
For instance, in some cases the users ``blindly'' delegate the decision to the AI without having access to their outcome and not being able to supervise it later~\cite{Chiang2021-kr,Maier2022-en,Zhang2020-wn,Fugener2022-fh}. 
\new{For example, in stock investment decisions, people can choose to invest directly in specific stocks or delegate a portion of their funds to the AI for future investment decisions~\cite{Maier2022-en}.}
In others, the AI agent has the decision to delegate and the user is assigned some of the decision making tasks to be completed on their own~\cite{Hemmer2023-hr,Fugener2022-fh}, or presented with the AI's outcome as support, resulting in the AI-first pattern~\cite{Bondi2022-ga}. In addition, the user can object to the delegation decision of the AI and take charge of the decision if considered appropriate~\cite{van2021moral}.

Lastly, the articles that we included in the ``Others'' category of interaction patterns can be separated into three groups. First, decision making tasks that involved more than one decision outcome~\cite{Porat2019-ak,M_A_Rahman2021-pk} and corresponding support from the AI agent. Second, decision making problems where multiple instances of decision tasks can take place and the interaction with the AI agent is continuous~\cite{Van_Berkel2022-uu,Fan2022-xt,Nourani2021-ji,Lindvall2021-lm,Reverberi_C2022-cc}. Third, interactions that involved a third agent~\cite{Wu2022-zy,Brachman2022-dq,banas2022}. In addition, a different case results when independent solutions, from the human and the AI agent, to the decision making problem are averaged as the final verdict~\cite{xiong2023partner}.

\subsection{Landscape over domains evaluated in AI-advised human decision making}
\paragraph{What domains have been defined as contexts to evaluate AI-assisted decision making?}
The selected articles included in this survey cover a broad range of different domain categories previously identified~\cite{lai2023towards}. Articles that included more than one experimental decision making task were counted toward more than one domain. Table \ref{tab:domains} presents a summary of the major domains and the different decision making tasks evaluated. Overall, the majority of the studies conducted human-AI interaction evaluations in real-world applications, with less than 15\% formulating artificial tasks. We included medical related databases in our search strategy, which contributes to the large representation of decision making tasks on the healthcare domain (26/108). In addition to healthcare, decision making tasks that may involve high-stakes outside of an experimental setup were identified in the finance and business (15/108), and law domains (6/108). 
The second most common domain was in the context of generic tasks (20/108) that are low-effort processing for humans but have mostly been used to develop AI benchmarks and demonstrate technical feasibility of algorithms. 
Other domains that typically include tasks targeted to non-expert users are social media (8), labeling (8) and leisure (1). We assigned tasks with unique applications to the Other domain (6/108). 
Even though we identified multiple decision making tasks in applications that require a specialized population, making it more difficult to achieve a large sample of participants, the majority of human-AI interactions have been evaluated with non-expert uses (60/108).   
The type of AI systems behind the interactions with users in the studies that covered these decision-making tasks were distributed among three categories: simulated models or Wizard of Oz experiments (39/105), deep learning-based models (34/105), and shallow models (35/105). 

\begin{table}[]
\caption{Domains and corresponding decision tasks used to study AI-assisted decision making.}
\label{tab:domains}
\begin{tabular}{  l   p{.6\linewidth}  p{.1\linewidth}  }
\textbf{Domain} & \textbf{Decision making task} & \textbf{Total tasks} \\ \hline
Education  & 
student performance prediction \cite{Rastogi2022-dh} 
& 1     \\ \hline
Artificial & 
identify the category of a shape \cite{Zhang2022-tx}; 
estimate quantities \cite{Hou2021-jk, Park2019-oo}; 
policy-verification task \cite{Nourani2021-ji}; 
quality control \cite{Yu2019-fi}; 
delivery method selection \cite{G_L_Liehner2022-ka}; 
pipe failure prediction \cite{Zhou2017-zl}; 
nutrition prediction \cite{Bucinca2020-ng}; 
object movement prediction \cite{Kumar2021-cm}; 
Memorizing images \cite{Allan2021-iv}; 
ranking \cite{Kim2023-ji}; 
spatial reasoning task \cite{Cao2022-tx}; 
pumping decisions \cite{xiong2023partner};
predict Titanic passenger's fate \cite{Baudel2021-ad}
& 14   \\ \hline
Finance/Business &  
stock market trading \cite{Cau2023-gq,Maier2022-en}; 
lending/loan assessment \cite{Jakubik2023-zv,Dikmen2022-yj,Appelganc2022-kv}; 
income prediction \cite{zhang2020effect,Alufaisan2021-yt}; 
revenue forecasting \cite{Vossing2022-si}; 
housing \cite{Prabhudesai2023-he,Tolmeijer2021-dr,Gupta2022-bf,Westphal2023-kn,Holstein2023-qj,Chiang2021-kr,Chiang2022-ix}
& 15 \\ \hline
Healthcare & 
Medical Diagnosis and Classification \cite{Gu2023-zl,Calisto2022-bx,Reverberi_C2022-cc,Reverberi_C2022-cc,Wang2022-jm,Schaekermann2020-eo,Hwang2022-dm,Tschandl2020-jv,Lam_Shin_Cheung2022-ho,Fogliato2022-uw,Van_Berkel2022-uu,Suresh2022-xp,Lindvall2021-lm,Gaube2023-xe,Cabitza2023-ik,Appelganc2022-kv};
Clinical Decision Support Systems and Treatment Planning \cite{van2021moral,Lee2021-lu,Jacobs2021-ic,Matthiesen2021-ee,Jiang2022-kt,Panigutti2022-va,Van_den_Brandt2020-fu,Porat2019-ak,Panigutti2022-va,Naiseh2023-ps,Bhattacharya2023-im}
& 26 \\ \hline
Generic& 
image classification \cite{Suresh2020-mx,Bondi2022-ga,Vodrahalli2022-fz,Fugener2022-fh,tejeda2022ai,Hemmer2023-hr,Gomez2023-la,Cabrera2023-kf};
text classification \cite{Smith-Renner2020-hs,Stites2021-om,Cabrera2023-kf,Cau2023-gq,Robbemond2022-ao,Riveiro2021-dv,Riveiro2022-qi,lai2020,Bansal2021-xt}; 
question answering \cite{Feng2022-ow, Silva2023-uy,Bansal2021-xt};
speech classification \cite{Tutul2021-bj,Zhang2022-ay}& 
20 \\ \hline
Labeling & 
text labeling \cite{Bernard2018-tx,Schrills2020-km,Ashktorab2021-us,Desmond2021-gw,Brachman2022-dq,Schemmer2023-ka,Mackeprang2019-dz}; image labeling \cite{Cau2023-gq}&
8 \\ \hline
Law &
recidivism prediction \cite{Grgic-Hlaca2019-cu,Wang2021-ra,Alufaisan2021-yt,Liu2021-gm} criminal referral decision \cite{Zytek2022-hr} penal sentence prediction \cite{Kahr2023-yh} &
6 \\ \hline
Leisure &
travel planning \cite{Khadpe2020-vy} &
1 \\ \hline
Social media &
friend matching \cite{M_A_Rahman2021-pk,Rechkemmer2022-yy}
content filtering \cite{M_A_Rahman2021-pk,Bunde2021-of,Lai2022-ye,Molina2022-ms}
fact checking \cite{Nguyen2018-ao,banas2022} &
8 \\ \hline
Professional &
human resources \cite{Peng2022-xt,Hofeditz2022-fq}; profession prediction \cite{Liu2021-gm} &
3  \\ \hline
Other &
environment \cite{Morrison2023-rr,Leichtmann2023-xx};
ethical decision-making \cite{Wu2022-zy,Tolmeijer2022-hm};
nutrition \cite{Bucinca2021-cv}
UX usability evaluation \cite{Fan2022-xt}&
6 \\ \hline
\end{tabular}
\end{table}

\begin{figure}[h]
    \centering
    \includegraphics[width=0.55\linewidth]{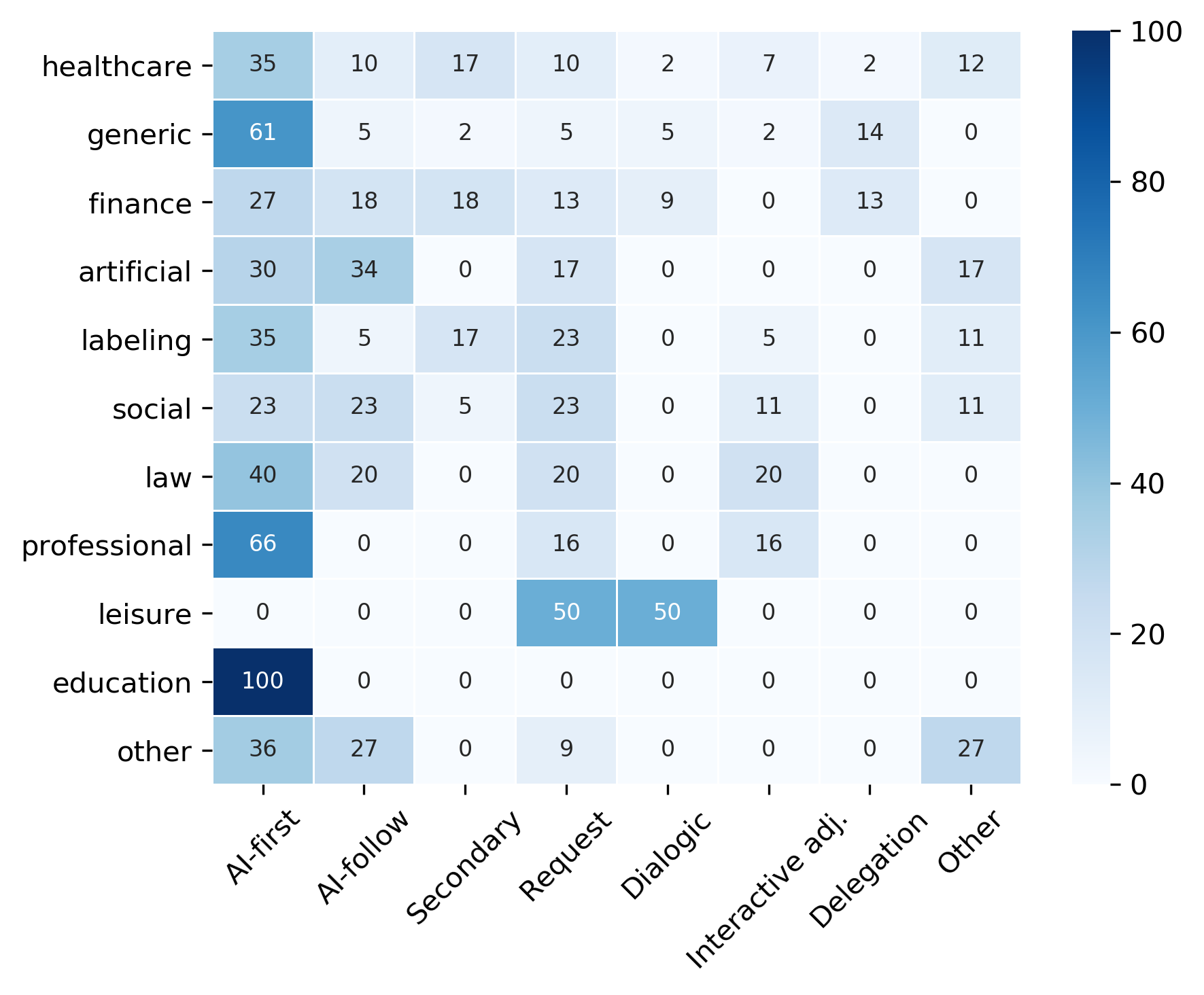}
    \caption{Percentage of interaction patterns observed in each domain of AI-assisted decision making tasks included in this survey. The numbers in the cells denote the percentage values (e.g., 17\% of the patterns identified in the healthcare domain correspond to Secondary assistance). One study can include multiple sequences of interaction and interaction patterns are not mutually exclusive. }
    \label{fig:domains}
\end{figure}

\paragraph{In what contexts were the interaction patterns observed during AI-assisted decision making processes?}
To better understand the existence and availability of the interaction patterns in different domains, we quantified the occurrence of patterns per domain and provide an overview describing the trends. 
Figure~\ref{fig:domains} shows the distribution of interaction patterns within our taxonomy for different domains. Values equal to zero mean that certain interaction pattern was not observed in the studies included in this survey for a specific domain.
Tables in the supplementary material provide more details on the interactions patterns for each paper included in this review.

Human-AI interactions in AI-advised decision makings in the healthcare domain mostly adhered to the AI-first assistance pattern (n=14), followed by Secondary assistance (n=7). Request-driven AI assistance was observed in a few cases (n=4) as well as AI-follow assistance (n=4). We identified one interaction that supported AI-guided dialogic user engagement, three interactions in the User-guided interactive adjustments, and one in which delegation was an option. Five sequences of interactions were in the ``Others'' category due to the higher complexity of the interactions. 
In the domain of finance and business, AI-first assistance was the most common type of interaction (n=6), followed by AI-follow assistance (n=4) and Secondary assistance (n=4). Some interactions supported Request-driven AI assistance (n=3), AI-guided dialogic user engagement (n=2), and delegation (n=3). We did no identify support for User-guided interactive adjustments. 
We identified four types of interaction patterns in the law and civic domain: AI-first assistance (n=4), AI-follow assistance (n=2), Request-driven assistance (n=2), and User-guided interactive adjustments (n=2). 
Decision-making tasks that involve professional related topics mostly followed AI-first assistance (n=4). Request-driven AI assistance and user-guided interactive adjustments were observed in one case each one. The other interaction patterns were not observed. 
Interaction patterns during decision making tasks involved in social media contexts were mostly of the AI-first (n=4), AI-follow (n=4), and Request-driven AI assistance types (n=4). We identified Secondary assistance in one case, and two cases in which User-guided interactive adjustments were enabled, in particular, for updating relevant terms for content moderation purposes. 
No Delegation or AI-guided dialogic user engagement was observed and two cases fell into the ``Others'' patterns category. 
During decision making tasks in generic applications, most of the interactions were dominated by the AI-first assistance (n=21), while only two cases involved the AI-follow assistance interaction. Regarding the more interactive patterns, two cases supported Request-driven AI assistance, two AI-guided dialogic user engagement, and one User-guided interactive adjustments. Only in one case the type of assistance was secondary. Delegation was featured in five cases. 
Labeling tasks mostly included AI-first assistance patterns (n=6), followed by Request-driven AI (n=4) and Secondary assistance (n=3), mostly clustering similar data points. 
We further identified AI-follow assistance and User-guided interactive adjustments once each one, and two types of interactions in the ``Others'' patterns category. 
Artificial decision making tasks were mostly dominated by AI-follow follow (n=8) and AI-first assistance (n=7). Request-driven AI assistance was featured in four cases and four interactions were in the ``Others'' patterns category. 
Lastly, AI-guided dialogic user engagement was the type of interaction in the leisure-related task and AI-first assistance in the task within the education domain. Interactions during decision-making tasks that belong to other domains mainly contained AI-first (n=4) and AI-follow (n=3) assistance types, or patterns in the ``Others'' category (n=3). Further, one case supported Request-driven AI assistance.

\paragraph{What evaluation methods and measures have been used in the study of AI-assisted decision making across domains?}
We describe the variety of evaluation measures used in the selected articles for our study of AI-assisted decision making. Table ~\ref{tab:metricsxdomain} in the Appendix presents a summary for each domain, separated by whether they correspond to objective or subjective measurements, and the constructs evaluated with the metrics. 
Broadly, we can observe a joint use of objective and subjective measures used in empirical evaluations of AI support for decision making tasks across multiple domains, as indicated by a previous review on interactive explanations~\cite{bertrand2023selective}. Objective measures covered four constructs overall. Efficacy is directly related to the joint (human and AI) task performance and is typically measured as the accuracy of decisions, errors, or other performance-related metrics. Trust and reliance construct is generally captured using agreement with AI advice or other variations (e.g., compliance frequency and over-reliance), weight of advice during AI-follow assistance, and delegation rate if the delegation pattern is present. Efficiency is considered during the execution of the decision-making task, directly measuring the actual decision time or more general total task time and exploration of other task functionalities. Although more commonly evaluated subjectively, questionnaires to capture AI model understanding have been developed~\cite{Wang2021-ra}. 

Some of the constructs captured through objective metrics can be evaluated from a subjective perspective as well, allowing researchers to identify potential mismatches between how users behaved and what they perceived during their interaction with AI. For instance, efficacy is easily quantifiable objectively, but has further been recorder from users' perceptions on their own, their AI partner's, or joint success to complete the task. Trust and reliance is another construct typically captured through standardized subjective questionnaires (e.g., ~\cite{korber2019theoretical,madsen2000measuring}) that cover multiple dimensions, such as competence, reliability, trustworthiness, intention to comply, credibility, uncertainty, among others. This last construct was evaluated in all the domains covered in this review but one (professional), highlighting the importance of developing trustworthy AI systems for humans to interact with. As with trust and reliance, the evaluation of system usability was covered in almost all the domains, except for education. Usability measurements include multiple dimensions such as usefulness, acceptance, satisfaction with the system, potential implementation, and system complexity. Another construct naturally captured through users' perception is decision satisfaction and mental demand, including ratings on confidence in the task, cognitive load, task difficulty or complexity, frustration, and workload. In order to evaluate users' processing of the information presented by the AI, understanding of these systems is commonly evaluated, especially when explainability components are involved. 
The construct of fairness did not stand out in the most frequent subjective measures but we attribute this to the fact that we excluded works in which decision making tasks were the context for evaluating humans' perceptions of the AI's outcomes, usually involving judgments of fairness. 
We included the metrics that did not fit into the categories mentioned above in a separate group labeled other.

\section{Discussion}
In establishing the interaction patterns presented in this paper, we drew from our observations of the Human-AI interactions used in AI-advised decision-making scenarios in prior empirical studies. While many of these studies effectively described their interfaces and user study procedures, there were instances where the information provided was not sufficient for a complete recreation of the interactions. Consequently, we had to carefully interpret and encode the interactions from these papers to the best of our abilities.

Constructing a taxonomy is inherently challenging due to the wide array of potential approaches that can be adopted. Specifically, in the domain of Human-AI interaction, interactions can be examined through various lenses (e.g., system-centric, oriented around user goal/task, distinguished by varying levels of granularity in interaction techniques). In this work, we have taken a preliminary step to structure an approach by integrating our perspectives
with observations of interaction paradigms used in existing studies. We hope this framework can serve as a foundation for further expansion and refinement to be made in subsequent research.

Using our taxonomy, we characterized existing interactions adopted in empirical studies. 
The most frequently observed interaction was the AI-first assistance pattern, where the user is given a direct presentation of the AI's prediction for the decision-making task. Its popularity can be attributed to the straightforwardness of demonstrating the effects of incorporating AI assistance into a decision-making task. However, it presents challenges, notably the difficulty in measuring the actual influence of AI assistance on user decision-making. Since the AI's solution is revealed before the user has had the opportunity to process the task independently, it can be convenient for the user to either dismiss or follow the AI's recommendation without sufficient reflection. 
\new{Noticeably, interaction patterns may engender different types of biases, and knowing them in advance may help guard against biases.} 
For instance, the AI-first interaction interaction also makes the user susceptible to the ``anchoring bias'', a phenomenon where a person's judgment is biased based on initial information. This bias can be avoided through the use of Secondary assistance, where the user must interpret supplementary information, determine its relevance, and decide how to incorporate it into their primary decision-making process.
In addition, direct presentation of AI inferences can lead to a lack of "sense of agency" for the user, which refers to the subjective feeling of controlling one's actions, and influencing external events through them ~\cite{wen2022sense}. Request-driven AI assistance can empower the user with the choice to view AI inferences and foster a sense of agency. It is worth noting that merely 10 of the identified AI-first assistance occurrences permitted the user to request AI inferences. However, although Request-driven AI assistance gives the user more sense of agency, it can give rise to possible confirmation bias or anchoring bias as well, especially when the user seeks explanations for decision verification or knowledge acquisition ~\cite{Barda_AJ2020-gy}. 

In contrast, in the AI-follow assistance pattern, the user is given a chance to solve the problem on their own, thus, potentially minimizing anchoring bias. Yet, whether users actually restart their decision making process is open to question. An article found that participants in this ``two-step'' workflow rarely revised their provisional diagnoses when the AI inferences differed from their earlier assessment~\cite{Fogliato2022-uw}. This hints at confirmation bias, a person's tendency to seek supporting evidence for their current hypothesis. In case the user does re-evaluate their prior assessments, the cognitive costs increase. Cognitive costs of re-examination, when new information becomes available, can be viewed as analogous to interruption and recovery on the initial task with new information~\cite{Fogliato2022-uw}. 
\new{Being the second most common pattern, the prevalence of the AI-follow pattern likely arises from a strong interest in disentangling the influence of AI advice on the human’s decision~\cite{vereschak2021evaluate}. Given their widespread adoption, understanding the pros and cons of AI-first and AI-follow approaches is crucial in developing AI systems that align with human cognitive processes and decision-making styles. Additionally, it is important to highlight the variability of data recorded regarding participants’ decisions with respect to the scenario at hand.}

A finer analysis per domain revealed the limited use and support for diverse interaction patterns, represented by most cell values equal to zero in Figure~\ref{fig:domains}. However, it is worth noticing that for high-stake domains, such as healthcare and finance, multiple interaction patterns have been explored when AI provides decision support.
\new{The choice of specific interaction patterns can be influenced by several factors including both design choices made by researchers and the intrinsic nature of the problems being addressed. In the former, the research intent and resources, including data availability and quality, may affect design choices. In the later, ethical and legal considerations play an important role, specially in high-stakes domains.
In more specialized fields, domain experts have their unique set of capabilities that can directly influence the choice and efficacy of interactions with the AI. For these experts, Secondary assistance can be beneficial, since they have the insights to effectively use the supplementary AI information for the primary decision-making task. However, for non-experts, the most suitable and beneficial choice of interaction pattern is unclear. Directly presenting a solution to the decision making problem can result in over-reliance~\cite{nourani2020role}, whereas Secondary assistance can avoid anchoring effects, but may not satisfy user needs. 
There remain questions about the universality of certain interaction patterns across varied user groups and task scenarios.} 

In general, while the most common patterns were AI-first, AI-follow, or Secondary assistance, in which the human role was limited to supervising the AI predictions, we did also note the presence of more dynamic interactions (e.g., Request-driven AI assistance, AI-guided dialogic user engagement, User-guided interactive adjustments), although less frequently. 
\new{Interactive elements can enhance the communication of explanations allowing users to interpret AI predictions through selection, mutation, and dialogue as suggested by~\cite{bertrand2023selective}. 
Likewise, supporting human input and review in the field of Interactive Machine Learning requires the design of interface elements for sample review, model inspection, or feedback assignment~\cite{dudley2018review}.  
Moving beyond supervising AI outputs, a different paradigm emerges when humans co-create solutions in partnership with AI systems, actively involving them in the decision-making process. 
By combining the strengths of each agent, human intuition and expertise synergize with AI's computational efficiency and data-driven insights for creating more robust solutions. 
Conversational user interfaces are central to enabling this rich interaction, which facilitates a two-way dialogue between humans and AI. 
Moreover, the concept of distributing decision-making responsibilities among different agents, as seen in delegation patterns, extends the assistance beyond individual decisions~\cite{Lai2022-ye}. 
This leads to diverse collaborative strategies, ranging from working in parallel---as exemplified in delegation scenarios where AI operates autonomously but in alignment with human intent---to more coordinated efforts, such as the turn-taking dynamic inherent in conversational AI. 
}

From our thorough exploration of interaction patterns in a sample of over 100 articles, we have identified gaps and opportunities. Much of the HCI literature on AI assistance has concentrated on intermittent scenarios (i.e., turn-taking). This is in contrast to continuous user interaction scenarios, where user input is sustained and can receive AI feedback at any given moment, which would be a more realistic and organic setup.
Another noticeable gap was regarding multi-agent teamwork collaboration. Most studies focus on interactions between a single user and a single AI, but there could be potential scenarios where an additional agent can be added. Also, while the current Delegation pattern allows the agent to change the assignment of decision-making, task coordination and allocation between agents can be explored. Moreover, we observed that the nature of the decision-making task is a notable point of consideration. A significant number of the interactions are evaluated within artificial tasks. Such controlled and simulated environments may not truly reflect the complexities faced in real-world applications. For instance, interactions in most studies end post-decisions, and decision-makers are not exposed to the full consequences of their choice \cite{kirkeboen2013revisions}. Thus, even when labeled as high-stakes, the lack of real consequences can influence user engagement. We must think more about differences in how users behave in experimental tasks and in equivalent real-life scenarios to assess whether AI assistance truly adds value to decision problems and capitalize on the findings of experimental evaluations.


The design choices for Human-AI interactions must be made with deliberation. Factors such as user psychology — including tendencies like confirmation bias and anchoring bias — user agency, and the cognitive costs in terms of time and effort, should all be taken into careful consideration.

\subsection{Limitations }
The works included in this survey are limited to published manuscripts that conduct empirical evaluations of human-AI interactions. This focus, while intentional, could have introduced certain limitations. The terminology used in the search could have excluded relevant work if the interaction design was not explicitly mentioned in the title or abstract, or even in the body of the text. Moreover, publication bias may have resulted in the exclusion of works relevant to this review. 
We also constrained our analysis to screen-based interfaces for AI-assistance, acknowledging that embodied AI might support additional interactions. Our strict selection criteria centered on studies encompassing complete decision-making tasks to ensure actual Human-AI interactions. Given the diversity of experimental designs and factors in the papers reviewed in this survey, we abstracted the interactions to discern patterns across the varied studies.

\section{Conclusion}
In this paper, we presented a systematic review of human-AI interactions in AI-advised decision making tasks that informed and grounded the formulation of a taxonomy of interaction patterns. 
Our proposed taxonomy of interaction patterns provides a structured foundation for understanding and designing these crucial interactions. It reveals that current practices often lean toward AI-driven or human-led decision processes, with limited emphasis on fostering interactive functionalities throughout the interactions. Recognizing the significance of interaction design, we advocate for deliberate choices in system development to enhance collaboration between humans and AI. Moving forward, the taxonomy presented here serves as a valuable resource to inform the design and development of AI-based decision support systems, ultimately fostering more productive, engaging, and user-centered collaborations.

\bibliographystyle{ACM-Reference-Format}
\bibliography{sample-base}

\appendix
\section*{Appendix}
\section{Search strategy}
\new{We used the following search string to screen titles, abstracts, and keywords in all the databases consulted:}

\new{(artificial intelligence OR machine learning OR deep learning OR neural network OR ai technology OR ai assistance OR ai OR algorithm OR hybrid intelligence) AND (human-ai collaboration OR human-ai interaction OR human-centered OR human centered OR human-ai teaming) AND (decision-making tasks OR ai decisions OR ai-assisted decision OR decision-making OR ai predictions OR ai-assisted decision-making OR ai-assisted diagnosis) AND (interaction OR interaction design OR interaction paradigm OR interface design)}

\new{We excluded articles with (``survey" OR ``workshop" OR ``guideline" OR "review") and (``robot" OR ``game" OR ``gaming" OR ``driving" OR ``autonomous vehicle" OR ``car" OR ``AV") in the title and abstract. }


\begin{figure}[h]
    \centering
    \includegraphics[width=0.90\linewidth]{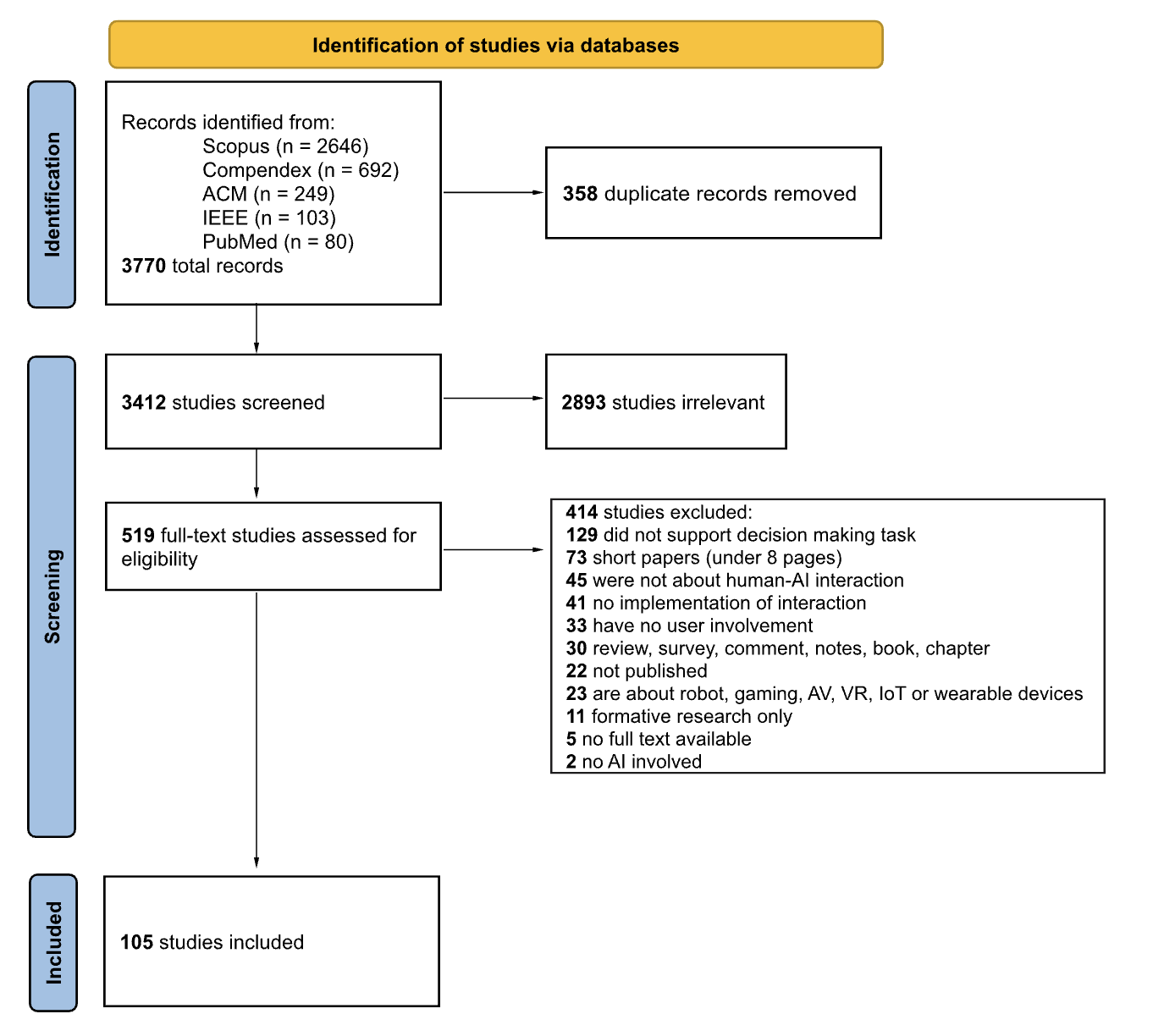}
    \caption{PRISMA diagram for article selection in this survey.}
    \label{fig:prisma}
\end{figure}

\section{Survey findings}
\begin{xltabular}{\textwidth}{>{\centering\arraybackslash}m{0.12\textwidth}|>{\centering\arraybackslash}m{0.09\textwidth}|>{\centering\arraybackslash}m{0.5\textwidth}|>{\centering\arraybackslash}m{0.1\textwidth}}

\caption{Metrics used to study AI-assisted decision making, grouped by domain. We classify metrics by the construct they were capturing and whether the correspond to objective or subjective measures.}
\label{tab:metricsxdomain}\\
Domain & \multicolumn{2}{c}{Metrics} & No. papers \\ \hline
\multirow{10}{*}{Artificial} & \multirow{3}{*}{Objective} & efficacy: decision accuracy
 & 5 \\ \cline{3-4} 
 &  & trust and reliance: user-AI agreement, delegation to AI rate, compliance frequency, gaze duration on AI
 & 10 \\ \cline{3-4} 
 &  & efficiency: decision time
 & 4 \\ \cline{2-4} 
 & \multirow{7}{*}{Subjective} & efficacy: perceived team performance, perceived expertise
 &  4 \\ \cline{3-4} 
 &  & trust and reliance: competence, predictability, dependability, responsibility, faith, reliability, trustworthiness, conformity, perceived expertise, perceived user-AI agreement
 & 8 \\ \cline{3-4} 
 &  & system usability: usefulness, helpfulness, pleasure
 & 3 \\ \cline{3-4} 
 &  & decision satisfaction \& mental demand: self-confidence, confidence in task, perceived AI's willingness to collaborate, cognitive load
 & 5 \\ \cline{3-4}
 &  & understanding: understanding of the system
 & 1 \\ \cline{3-4}
 &  & fairness: responsibility attribution of loss
 & 1 \\ \cline{3-4}
 &  & other: decision risk perception
 & 1 \\\hline
 \multirow{3}{*}{Education} & \multirow{2}{*}{Objective} & efficacy: decision accuracy
 & 1 \\ \cline{3-4} 
 &  & trust and reliance: user-AI agreement
 & 1 \\ \cline{2-4} 
 & Subjective & trust and reliance: reliability, trustworthiness, intention to comply
 & 1 \\\hline
\multirow{9}{*}{Finance/Business} & \multirow{3}{*}{Objective} & efficacy: decision accuracy, errors, performance
 & 11 \\ \cline{3-4} 
 &  & trust and reliance: user-AI agreement, delegation to AI rate, weight of advice, decision outcome, overreliance, compliance frequency, number of options explored by user, decision time
 & 13 \\ \cline{3-4} 
 &  & efficiency: task duration, number of options explored by user, decision time
 & 3 \\ \cline{2-4} 
 & \multirow{6}{*}{Subjective} & efficacy: perceived self performance, perceived model performance
 &  2 \\ \cline{3-4} 
 &  & trust and reliance: reliability, trustworthiness, intention to comply
 & 8 \\ \cline{3-4} 
 &  & system usability: usefulness, enjoyment, satisfaction
 & 3 \\ \cline{3-4} 
 &  & decision satisfaction \& mental demand: confidence in task, task difficulty, task complexity
 & 3 \\ \cline{3-4}
 &  & understanding: understanding of the system, understanding of the tutorial
 & 2 \\ \cline{3-4}
 &  & other: decision strategies, quality of explanations, perception of uncertainty information, interpretation and incorporation of uncertainty information into final recommendations
 & 3 \\\hline
  \multirow{9}{*}{Generic} & \multirow{3}{*}{Objective} & efficacy: decision accuracy
 & 16 \\ \cline{3-4} 
 &  & trust and reliance: user-AI agreement, delegation to AI rate, weight of advice, overtrust, undertrust, compliance frequency, decision time
 & 13 \\ \cline{3-4} 
 &  & efficiency: task duration, decision time
 & 4 \\ \cline{2-4} 
 & \multirow{8}{*}{Subjective} & efficacy: perceived model accuracy, perceived task performance, perceived model capability, perceived self performance
 &  2 \\ \cline{3-4} 
 &  & trust and reliance: reliability, trustworthiness, perceived AI uncertainty, perceived credibility, perceived agreement
 & 9 \\ \cline{3-4} 
 &  & system usability: usefulness, satisfaction, helpfulness, likeliness to use again
 & 7 \\ \cline{3-4} 
 &  & decision satisfaction \& mental demand: confidence in task, frustration
 & 4 \\ \cline{3-4}
 &  & understanding: understanding of the system
 & 3 \\ \cline{3-4}
 &  & efficiency: perceived efficiency, perceived effectiveness
 & 1 \\ \cline{3-4}
 &  & fairness: perceived AI contribution, perceived teamwork
 & 1 \\ \cline{3-4}
 &  & other: quality of explanations, feedback importance, helpfulness of visualization, suggestions
 & 6 \\\hline
 \multirow{11}{*}{Healthcare} & \multirow{3}{*}{Objective} & efficacy: decision accuracy, specificity, sensitivity, coefficient of variability, inter-rater reliability
 & 17 \\ \cline{3-4} 
 &  & trust and reliance: user-AI agreement, delegation to AI rate, weight of advice, overtrust, compliance frequency, decision time, appropriate trust, dominance
 & 10 \\ \cline{3-4} 
 &  & efficiency: task duration, decision time, number of tasks completed
 & 7 \\ \cline{3-4} 
 &  & other: user epistemic uncertainty
 & 1 \\ \cline{2-4} 
 & \multirow{8}{*}{Subjective} & efficacy: perceived model accuracy
 &  3 \\ \cline{3-4} 
 &  & trust and reliance: reliability, trustworthiness, integrability perception
 & 12 \\ \cline{3-4} 
 &  & system usability: usefulness, satisfaction, perceived acceptance, possibility of adoption/implementation, perceived hindrance due to AI, richness, perceived utility
 & 11 \\ \cline{3-4} 
 &  & decision satisfaction \& mental demand: confidence in task, confidence in AI, mental support, workload, task difficulty
 & 11 \\ \cline{3-4}
 &  & understanding: understanding of the system, explainability, predictability
 & 4 \\ \cline{3-4}
 &  & efficiency: perceived efficiency
 & 2 \\ \cline{3-4}
 &  & fairness: moral compliance, feeling of control
 & 2 \\ \cline{3-4}
 &  & other: quality and usefulness of explanations, perception free text
 & 5  \\\hline
  \multirow{11}{*}{Labeling} & \multirow{3}{*}{Objective} & efficacy: decision accuracy, F-1 score, precision, recall
 & 7 \\ \cline{3-4} 
 &  & trust and reliance: user-AI agreement, overreliance, decision time, delegation to AI rate
 & 5 \\ \cline{3-4} 
 &  & efficiency: task duration, decision time, number of tasks completed
 & 5 \\ \cline{2-4} 
 & \multirow{8}{*}{Subjective} & trust and reliance: reliance, perceived AI uncertainty, trustworthiness, perceived agreement
 & 3 \\ \cline{3-4} 
 &  & system usability: usefulness, acceptance, usability
 & 3 \\ \cline{3-4} 
 &  & decision satisfaction \& mental demand: confidence in task, self-confidence, satisfaction
 & 4 \\ \cline{3-4}
 &  & understanding: observability
 & 1 \\ \cline{3-4}
 &  & fairness: perceived AI contribution
 & 5 \\\hline
  \multirow{9}{*}{Law} & \multirow{4}{*}{Objective} & efficacy: decision accuracy
 & 3 \\ \cline{3-4} 
 &  & trust and reliance: decision time, compliance frequency, appropriate trust, overtrust, undertrust, user-AI agreement, weight on advice
 & 5 \\ \cline{3-4} 
 &  & efficiency: decision time
 & 1 \\ \cline{3-4} 
 &  & understanding: objective understanding
 & 1 \\ \cline{2-4} 
 & \multirow{5}{*}{Subjective} & trust and reliance: trustworthiness
 & 2 \\ \cline{3-4} 
 &  & system usability: usefulness, helpfulness
 & 2 \\ \cline{3-4} 
 &  & decision satisfaction \& mental demand: confidence in task
 & 3 \\ \cline{3-4}
 &  & understanding: subjective understanding
 & 1 \\ \cline{3-4}
 &  & other: self-reported decision-making strategies
 & 1 \\\hline
 \multirow{2}{*}{Leisure} & {Objective} & efficiency: duration of interaction, word count
 & 1 \\ \cline{2-4} 
 & {Subjective} & system usability: perceived usability and warmth, intention to adopt, desire to cooperate
 & 1 \\\hline
 \multirow{4}{*}{Professional} & \multirow{2}{*}{Objective} & efficacy: decision accuracy, true positive rate
 & 2 \\ \cline{3-4} 
 &  & trust and reliance: user-AI agreement
 & 2 \\ \cline{2-4} 
 & \multirow{1}{*}{Subjective} & system usability: usefulness
 & 1 \\\hline
 \multirow{7}{*}{Social Media} & \multirow{2}{*}{Objective} & efficacy: decision accuracy and union precision, true and false positive difference 
 & 2 \\ \cline{3-4} 
 &  & trust and reliance: user-AI agreement, compliance frequency
 & 2 \\ \cline{2-4}
 & \multirow{5}{*}{Subjective} & trust and reliance: observed security, reliability, faith, perceived agreement, trustworthiness
 & 5 \\ \cline{3-4} 
 & & efficacy: perceived model accuracy \& competence
 & 3 \\ \cline{3-4} 
 & & system usability: usefulness, ease of use, intention to use
 & 1 \\ \cline{3-4} 
 &  & decision satisfaction and mental demand: confidence in task, workload
 & 1 \\ \cline{3-4}
 &  & understanding: understanding of the system
 & 2 \\\hline
 \multirow{7}{*}{Other} & \multirow{2}{*}{Objective} & efficacy: decision accuracy, precision, recall
 & 3 \\ \cline{3-4} 
 &  & trust and reliance: user-AI agreement, compliance frequency, overreliance
 & 5 \\ \cline{2-4} 
 & \multirow{5}{*}{Subjective} & trust and reliance: trustworthiness, perceived credibility
 & 5 \\ \cline{3-4} 
 & & system usability: helpfulness, system complexity, satisfaction
 & 3 \\ \cline{3-4} 
 &  & decision satisfaction \& mental demand: confidence in task, mental demand
 & 2 \\ \cline{3-4} 
 &  & fairness: responsibility attribution
 & 1 \\ \cline{3-4} 
 &  & understanding: understanding of the system
 & 2 \\\hline
\end{xltabular}

\end{document}


\title[Human-AI collaboration is not very collaborative yet]{Human-AI collaboration is not very collaborative yet: A taxonomy of interaction patterns in AI-assisted decision making from a systematic review}

\maketitle

\section*{Supplementary material}

\begin{xltabular}{\textwidth}{>{\centering\arraybackslash}m{0.05\textwidth}|>{\centering\arraybackslash}m{0.14\textwidth}|>{\centering\arraybackslash}m{0.14\textwidth}|>{\centering\arraybackslash}m{0.12\textwidth}|>{\centering\arraybackslash}m{0.13\textwidth}|>{\centering\arraybackslash}m{0.13\textwidth}|>{\centering\arraybackslash}m{0.10\textwidth}}

\caption{A summary of the context-related information extracted from the selected articles.}
\label{tab:extraction_context}\\
\textbf{Ref.}&\textbf{Domain}& \textbf{Decision task}&\textbf{Task difficulty}&\textbf{Task stimulus}&\textbf{Type of AI}&\textbf{AI outcome}\\
\hline
        \cite{Zhang2022-tx}&   Artificial    &  shape classification   &   Superhuman   &   Images    &    mock-up     &    shape class   \\ \hline
        \cite{Xiong2023-iw} & Artificial & pumping balloons & no expertise & features and possible profits & mock-up & pumping choice \\ \hline
        \cite{Cao2022-tx} & Artificial & spatial reasoning task & no expertise & images & mock-up & 2D image slice of 3d structures\\ \hline
        \cite{Kim2023-ji} & Artificial & item ranking & no expertise & features & mock-up & items ranking  \\ \hline
        \cite{Allan2021-iv} & Artificial & image memorization & no expertise & images & mock-up & image contents  \\ \hline
        \cite{Kumar2021-cm} & Artificial & object movement prediction & no expertise & images & mock-up & movement: left or right  \\ \hline
        \cite{Baudel2021-ad} & Artificial & predict Titanic passenger’s fate & no expertise & features & shallow models & binary: survival or not  \\ \hline
        \cite{Zhou2017-zl} & Artificial & pipe failure prediction & requires expertise & pipe features & mock-up & pipe failure rate  \\ \hline
        \cite{G_L_Liehner2022-ka} & Artificial & delivery method selection & no expertise & context and risk features & mock-up & automatic/manual delivery  \\ \hline
        \cite{Yu2019-fi} & Artificial & quality control & no expertise & images & mock-up & good/faulty glass  \\ \hline
        \cite{Park2019-oo} & Artificial & quantity estimation & no expertise & images & mock-up & quantity \\ \hline
        \cite{Nourani2021-ji} & Artificial & policy-verification & no expertise & video frames & DL model & presence of actions \\ \hline
        \cite{Hou2021-jk} & Artificial & quantity estimation & no expertise & images & mock-up & quantity\\ \hline
        \cite{Bucinca2020-ng} & Artificial & nutrition prediction & no expertise & images & mock-up & percent fat threshold  \\ \hline
        \cite{Rastogi2022-dh} & Education & student performance prediction & no expertise & features & shallow models & student performance \\ \hline
        \cite{Prabhudesai2023-he} & Finance/Business & house price estimation & requires expertise & housing features & shallow models & house selling price  \\ \hline
        \cite{Westphal2023-kn} & Finance/Business & hotel room price estimation & requires expertise & features  & shallow models & room price  \\ \hline
        \cite{Jakubik2023-zv} & Finance/Business & lending/loan assessment & no expertise & features  & shallow models & borrower decision and outcome  \\ \hline
        \cite{Vossing2022-si} & Finance/Business & revenue forecasting & requires expertise & features, time-series data & DL models & restaurant's revenue  \\ \hline
        \cite{Maier2022-en} & Finance/Business & investment & no expertise & features  & mock-up & amount of investment  \\ \hline
        \cite{Dikmen2022-yj} & Finance/Business & lending/loan assessment & requires expertise & features  & shallow models & predicted probabilities  \\ \hline
        \cite{Chiang2022-ix} & Finance/Business & house price estimation & no expertise & housing features & shallow models & house selling price  \\ \hline
        \cite{Tolmeijer2021-dr} & Finance/business & house search & no expertise & housing features & mock-up & house candidates  \\ \hline
        \cite{Chiang2021-kr} & Finance/Business & house price estimation & no expertise & housing features & shallow models & house selling price  \\ \hline
        \cite{Gupta2022-bf} & Finance/business & house search & no expertise & housing features & mock-up & house candidates  \\ \hline
        \cite{zhang2020effect} & Finance/Business & income prediction & no expertise & features  & shallow models & income prediction (binary)  \\ \hline
        \cite{Holstein2023-qj} & Finance/Business & house price estimation & no expertise & housing features & shallow models & house selling price  \\ \hline
        \cite{Cau2023-gq} & Finance/Business & stock market trading & requires expertise & historical stock price & shallow models & buy/sell action\\ \hline
        \cite{Gomez2023-la} & Generic & image classification & requires expertise & images & mock-up & bird category  \\ \hline
        \cite{Vodrahalli2022-fz}  & Generic & image classification & no expertise & images & mock-up & binary class \\ \hline
        \cite{Hemmer2023-hr} & Generic & image classification & no expertise & images & DL models & image class  \\ \hline
        \cite{tejeda2022ai} & Generic & image classification & no expertise & images & DL models & image class  \\ \hline
        \cite{Silva2023-uy} & Generic & question answering & no expertise & text & mock-up & answer choice  \\ \hline
        \cite{Bondi2022-ga} & Generic & image classification & no expertise & images & shallow models & animal present or not  \\ \hline
        \cite{Fugener2022-fh} & Generic & image classification & no expertise & images & DL models & image class  \\ \hline
        \cite{Zhang2022-ay} & Generic & image classification: predict emotion  & no expertise & 2D image (spectrogram) & DL models & emotion class   \\ \hline
        \cite{Feng2022-ow} & Generic & question answering & no expertise & text & DL models & response option  \\ \hline
        \cite{Tutul2021-bj} & Generic & speech classification & requires expertise & acoustic features from speech signals  & DL models & public speaking anxiety levels  \\ \hline
        \cite{Riveiro2021-dv} & Generic & text classification & no expertise & text & shallow models & text class  \\ \hline
        \cite{Stites2021-om} & Generic & text classification & no expertise & text features & shallow models & benign/spam   \\ \hline
        \cite{Suresh2020-mx} & Generic & image classification & requires expertise & images & DL models & image class  \\ \hline
        \cite{lai2020} & Generic & text classification & superhuman task & text & shallow models, DL models & deceptive/genuine   \\ \hline
        \cite{Smith-Renner2020-hs} & Generic & text classification & no expertise & unigram features & DL models & email classification  \\ \hline
        \cite{Riveiro2022-qi} & Generic & text classification & no expertise & text & mock-up & text class  \\ \hline
        \cite{Robbemond2022-ao} & Generic & text classification & no expertise & text & mock-up & credible/not credible \\ \hline
        \cite{Gaube2023-xe} & Healthcare & x-ray diagnosis & requires expertise & x-ray images & mock-up & diagnosis \\ \hline
        \cite{Bhattacharya2023-im} & Healthcare & risk of diabetes prediction & no expertise & tabular data & shallow models & risk of disease, risk recovery, recommendations to reduce risk \\ \hline
        \cite{Cabitza2023-ik} & Healthcare & ECG analysis & requires expertise & ECG & mock-up & diagnosis \\ \hline
        \cite{Naiseh2023-ps} & Healthcare & prescriptions screening & requires expertise & patient features & mock-up & accept/reject a prescription \\ \hline
        \cite{Du2022-od} & Healthcare & diabetes diagnosis & requires expertise & patient features & shallow models & risk disease (low vs high) \\ \hline
        \cite{Jiang2022-kt} & Healthcare & telemedicine self-diagnosis & requires expertise & patient features & shallow models & possible diagnosis \\ \hline
        \cite{Fogliato2022-uw} & Healthcare & x-ray diagnosis & requires expertise & x-ray image & DL models & diagnosis \\ \hline
        \cite{Panigutti2022-va} & Healthcare & acute myocardial infarction diagnosis & requires expertise & patient features & DL models & diagnosis \\ \hline
        \cite{Calisto2022-bx} & Healthcare & breast cancer diagnosis  & requires expertise & images & DL models & severity classification \\ \hline
        \cite{Van_Berkel2022-uu} & Healthcare & polyp identification  & requires expertise & endoscopy video & mock-up  & areas flagged as polyps \\ \hline
        \cite{Suresh2022-xp} & Healthcare & electrocardiogram beat classification & requires expertise & time series & DL models & normal/abnormal \\ \hline
        \cite{Lam_Shin_Cheung2022-ho} & Healthcare & chest x-ray classification & requires expertise & chest x-rays & DL models & normal/abnormal \\ \hline
        \cite{Hwang2022-dm} & Healthcare & sleep stage classification & requires expertise & EEG & DL models & sleep stages \\ \hline
        \cite{Matthiesen2021-ee} & Healthcare & clinical action & requires expertise & time series data, features & shallow models & risk of VT/VF within 30 days \\ \hline
        \cite{Van_der_Waa2021-fz} & Healthcare & medical triage task  & requires expertise & features & shallow models & IC-unit/hospital ward/home \\ \hline
        \cite{Lee2021-lu} & Healthcare & rehabilitation exercise assessment & requires expertise & sensor data (videos and features) & DL models & quality of rehabilitation \\ \hline
        \cite{Lindvall2021-lm} & Healthcare & metastasis assessment & requires expertise & images & DL models & number of positive and negative lymph nodes \\ \hline
        \cite{Jacobs2021-ic} & Healthcare & treatment selection & requires expertise & features & mock-up & treatment selection \\ \hline
        \cite{Tschandl2020-jv} & Healthcare	& skin cancer diagnosis	& requires expertise & skin images & DL models & diagnostic category \\ \hline
        \cite{Schaekermann2020-eo} & Healthcare & sleep stage classification & requires expertise & time series data - polysomnograms & DL models & sleep stages \\ \hline
        \cite{Van_den_Brandt2020-fu} & Healthcare & glaucoma progression prediction & requires expertise & OCT scans & DL models & visual field measurements \\ \hline
        \cite{Porat2019-ak} & Healthcare & treatment selection & requires expertise & patient features & shallow models & recurrent stroke risk, treatment options \\ \hline
        \cite{Wang2022-jm} & Healthcare & brain disease classification & requires expertise & CT volumes & DL models & diagnosis \\ \hline
        \cite{Reverberi_C2022-cc} & Healthcare & colorectal lesions diagnosis & requires expertise & videos & DL models & adenoma or not \\ \hline
        \cite{Jungmann_F2023-lk} & Healthcare & breast lesion classification & requires expertise & mammograms & DL models & malignant/benign \\ \hline
        \cite{Gu2023-zl} & Healthcare & meningioma grading & requires expertise & images & DL models & grade level \\ \hline
        \cite{Schemmer2023-ka} & Labeling & deceptive/genuine hotel review & no expertise & text & shallow models & binary label \\ \hline
        \cite{Brachman2022-dq} & Labeling & labeling conflict resolution & no expertise & text & shallow models & text answer \\ \hline
        \cite{Desmond2021-gw} & Labeling & data labeling & no expertise & text & shallow models & customer intent class \\ \hline
        \cite{Ashktorab2021-us} & Labeling & labeling for text classification & no expertise & text & shallow models & intent class \\ \hline
        \cite{Schrills2020-km} & Labeling & character classification & no expertise & images & DL models & number class \\ \hline
        \cite{Mackeprang2019-dz} & Labeling & annotation of concept ideas  & no expertise & text & shallow models & concepts \\ \hline
        \cite{Bernard2018-tx} & Labeling & visual-interactive labeling (VIL) & no expertise
        & images & shallow models & integer class \\ \hline
        \cite{Zytek2022-hr} & Law & criminal referral decision & requires expertise & features  & shallow models & risk score \\ \hline
        \cite{Wang2021-ra} & Law & recidivism prediction & requires expertise & features  & shallow models & binary \\ \hline
        \cite{Grgic-Hlaca2019-cu} & Law & recidivism prediction & requires expertise & features  & shallow models & binary \\ \hline
        \cite{Kahr2023-yh} & Law & penal sentence prediction & requires expertise & text  & mock-up & jail time \\ \hline
        \cite{Khadpe2020-vy} & Leisure & travel planning & no expertise & text & mock-up & hotel and flight choice \\ \hline
        \cite{Leichtmann2023-xx} & Other: environment & mushroom poisonous or not & requires expertise & images & DL models & edible/poisonous \\ \hline
        \cite{Morrison2023-rr} & Other: environment & assessing building damage & requires expertise & satellite images & mock-up & damage assessment \\ \hline
        \cite{Tolmeijer2022-hm} & Other: Ethics & operation decisions & no expertise & features & mock-up & location \\ \hline
        \cite{Wu2022-zy} & Other: Ethics & moral dilemma & no expertise & features & mock-up & moral dilemma outcome \\ \hline
        \cite{Bucinca2021-cv} & Other: Nutrition & diet arrangements & no expertise & image & mock-up & healthy ingredient replacements \\ \hline
        \cite{Fan2022-xt} & Other: UX & UX usability evaluation & requires expertise & video & mock-up & usability problem \\ \hline
        \cite{Hofeditz2022-fq} & Professional & screening job applicants & requires expertise & features & mock-up & group of similar applications with the strongest qualifications \\ \hline
        \cite{Peng2022-xt} & Professional & screening job applicants & no expertise & text & DL models & match or not \\ \hline
        \cite{banas2022} & Social Media & fact checking & requires expertise & text & mock-up & true/false \\ \hline
        \cite{Molina2022-ms} & Social Media & content filtering & no expertise & text & mock-up & hate/not hate speech; suicidal ideation/not suicidal \\ \hline
        \cite{Rechkemmer2022-yy} & Social Media & dating events prediction & no expertise & features & mock-up & binary \\ \hline
        \cite{Lai2022-ye} & Social Media & content filtering & no expertise & text & DL models & toxic/non toxic  \\ \hline
        \cite{Lu2021-xi} & Social Media & dating events prediction & no expertise & features & mock-up & binary \\ \hline
        \cite{Bunde2021-of} & Social Media & content filtering & no expertise & text & DL models & hate speech/not hate speech \\ \hline
        \cite{Nguyen2018-ao} & Social Media & fact checking & no expertise & text & shallow models & true/false \\ \hline
        \cite{M_A_Rahman2021-pk} & Social Media & friend matching, content filtering & superhuman task & features & shallow models, DL models & binary \\ \hline
        \cite{Appelganc2022-kv} & Two: finance \& healthcare & loan assignment, x-ray diagnosis & no expertise & features, images & mock-up & loan acceptance, percentage of malignant tissue \\ \hline
        \cite{Cabrera2023-kf} & Two: image classification, text classification  & fake review detection, image classification & no expertise & images, text & mock-up & fake/true review, satellite image class, bird image class \\ \hline
        \cite{Cau2023-nc} & Two: labeling and text classification & digit identification, text classification & no expertise & image, text & DL models & digit label, positive/negative \\ \hline
        \cite{Alufaisan2021-yt} & Two: law and finance  & recidivism prediction, income prediction & no expertise & features & shallow models & recidivism prediction, low/high income \\ \hline
        \cite{Liu2021-gm} & Two: Law and profession & recidivism prediction, profession prediction & no expertise & text & shallow models & recidivism prediction, profession prediction: multi-class \\ \hline
        \cite{Bansal2021-xt} & Two: text classification and question answering  & text classification, question answering  & no expertise & text & DL models & positive/negative sentiment analysis, correct/incorrect answer \\ \hline
\end{xltabular}

\begin{xltabular}{\textwidth}{>{\centering\arraybackslash}m{0.05\textwidth}|>{\centering\arraybackslash}m{0.13\textwidth}|>{\centering\arraybackslash}m{0.14\textwidth}|>{\centering\arraybackslash}m{0.13\textwidth}|>{\centering\arraybackslash}m{0.13\textwidth}|>{\centering\arraybackslash}m{0.13\textwidth}|>{\centering\arraybackslash}m{0.08\textwidth}}
\caption{A summary of the information extracted from the selected articles regarding the empirical study. AI presentations corresponds to the way the AI system was presented to the users. The Objective and Subjective columns comprise the corresponding evaluation measures. AI and User actions are specified using the interaction building blocks (the output in the action-output pair was omitted for clarity in the table). The interaction patterns from our taxonomy are in the Patterns column and shortened for clarity.}
\label{tab:extraction_interactions}\\
\textbf{Ref.}& \textbf{AI presentation}&\textbf{Objective}&\textbf{Subjective}&\textbf{AI actions}&\textbf{User actions} & \textbf{Patterns}\\
\hline
\cite{Zhang2022-tx}  &     ShapeBot    &  Efficacy, trust and reliance   &     trust and reliance    &    predict, display    &   predict, decide & AI-follow\\
\hline
\cite{Xiong2023-iw} & machine & efficacy, trust and reliance & trust and reliance, efficacy, decision satisfaction \& mental demand, fairness, system usability & predict & predict, decide & AI-follow, Others \\ \hline
        \cite{Vodrahalli2022-fz} & AI & trust and reliance, efficacy & NA & predict, display & predict, decide & AI-follow \\ \hline
        \cite{Cao2022-tx} & AI & efficacy, trust and reliance & trust and reliance, decision satisfaction \& mental demand & predict & predict, decide & AI-follow \\ \hline
        \cite{Kim2023-ji} & AI & trust and reliance, efficiency & trust and reliance & predict, display & predict, decide & AI-follow \\ \hline
        \cite{Allan2021-iv} & AI & efficacy & trust and reliance & predict, display & decide & AI-first \\ \hline
        \cite{Kumar2021-cm} & computer & trust and reliance & decision satisfaction \& mental demand & predict, display & predict, request, decide & AI-follow, Request-driven \\ \hline
        \cite{Baudel2021-ad} & algorithm & trust and reliance, efficiency & efficacy & predict & decide, request & AI-first, Request-driven \\ \hline
        \cite{Zhou2017-zl} & Automatic Predictive Assistant & NA & trust and reliance, decision satisfaction \& mental demand & predict, display & decide & AI-first \\ \hline
        \cite{G_L_Liehner2022-ka} & artificial intelligence & trust and reliance, efficiency & decision risk perception & predict, display & decide & AI-first \\ \hline
        \cite{Yu2019-fi} & automatic quality monitor  & efficacy, trust and reliance & efficacy, trust and reliance & predict & decide & AI-first \\ \hline
        \cite{Park2019-oo} & machine vision algorithm & trust and reliance & trust and reliance & predict & predict, request, decide & AI-follow, Request-driven \\ \hline
        \cite{Nourani2021-ji} & system & NA & efficacy, system usability & provide, display & request, predict & Others \\ \hline
        \cite{Hou2021-jk} & AI & trust and reliance & system usability & predict & predict, decide & AI-follow \\ \hline
        \cite{Bucinca2020-ng} & AI & efficacy & understanding, trust and reliance, system usability & predict, display & decide & AI-first \\ \hline
        \cite{Rastogi2022-dh} & AI & trust and reliance, efficiency & trust and reliance & predict, display & decide & AI-first \\ \hline
        \cite{Prabhudesai2023-he} & algorithm & NA & trust and reliance, perception of uncertainty information & provide, display & predict, request & Secondary, Request-driven \\ \hline
        \cite{Westphal2023-kn} & machine learning model & trust and reliance & trust and reliance, understanding, decision satisfaction \& mental demand & predict, display & predict, decide & AI-follow \\ \hline
        \cite{Jakubik2023-zv} & AI & efficacy, trust and reliance & NA & predict, display & decide & AI-first \\ \hline
        \cite{Vossing2022-si} & AI system & efficacy, trust and reliance & trust and reliance & predict, display & decide, request & AI-first, Request-driven \\ \hline
        \cite{Maier2022-en} & AI & efficacy, trust and reliance & NA & predict, delegate & predict, delegate & Delegate \\ \hline
        \cite{Dikmen2022-yj} & machine learning assistant & trust and reliance & trust and reliance, decision satisfaction \& mental demand, quality of explanations & provide, display & predict & Secondary \\ \hline
        \cite{Chiang2022-ix} & ML model & efficacy, trust and reliance & understanding of the tutorial, usefulness, enjoyment, satisfaction & predict & predict, decide & AI-follow \\ \hline
        \cite{Tolmeijer2021-dr} & intelligent system & efficacy, trust and reliance & trust and reliance & predict & predict, request, decide & AI-follow, Request-driven \\ \hline
        \cite{Chiang2021-kr} & machine learning model & efficacy, trust and reliance & efficacy & predict & predict, delegate & Delegate \\ \hline
        \cite{Gupta2022-bf} & Chat & trust and reliance, efficiency, efficacy & trust and reliance, decision satisfaction \& mental demand, quality of explanations & request, predict & collect, decide & AI-guided dial. \\ \hline
        \cite{zhang2020effect} & AI & efficacy, trust and reliance & NA & predict, display & predict, decide, delegate & AI-follow, Delegate \\ \hline
        \cite{Holstein2023-qj} & model & efficacy & system usability & predict, display & decide & AI-first \\ \hline
        \cite{Cau2023-gq} & AI & efficacy, trust and reliance & trust and reliance & predict, display & decide & AI-first \\ \hline
        \cite{Gomez2023-la} & AI assistant & trust and reliance, efficiency & trust and reliance, system usability, fairness, suggestions & predict, request, provide  & request, decide, collect & AI-first, Request-drive, AI-guided dial. \\ \hline
        \cite{Hemmer2023-hr} & AI  & efficacy & efficacy, system usability & predict, delegate, display & predict & Delegate \\ \hline
        \cite{tejeda2022ai} & AI & efficacy & decision satisfaction \& mental demand & predict, display & predict, decide & AI-follow, AI-first \\ \hline
        \cite{Silva2023-uy} & robot & efficacy, trust and reliance, efficiency & trust and reliance, understanding, perception of explanations & predict, display & decide & AI-first \\ \hline
        \cite{Bondi2022-ga} & AI Model & efficacy, trust and reliance & NA & delegate, predict & predict, decide & Delegate \\ \hline
        \cite{Fugener2022-fh} & AI & efficacy, trust and reliance & efficacy & predict, delegate & predict, delegate & Delegate \\ \hline
        \cite{Zhang2022-ay} & system & efficacy, trust and reliance & trust and reliance, system usability, decision satisfaction \& mental demand & predict, display & decide & AI-first \\ \hline
        \cite{Feng2022-ow} & AI & efficacy & NA & predict, display, provide, modify & decide & AI-first \\ \hline
        \cite{Tutul2021-bj} & machine learning algorithm  & efficacy, trust and reliance & trust and reliance & predict, display & decide & AI-first \\ \hline
        \cite{Riveiro2021-dv} & AI system & trust and reliance & satisfaction with explanations, understanding, efficacy & predict, display & decide & AI-first \\ \hline
        \cite{Stites2021-om} & model & efficacy, trust and reliance, efficiency & trust, helpfulness of visualization, perceived performance (frequency of correct responses) & predict, display & decide & AI-first \\ \hline
        \cite{Suresh2020-mx} & model, machine learning & efficacy, trust and reliance & NA & predict, display & decide, request & AI-first, Request-driven \\ \hline
        \cite{lai2020} & AI & efficacy & NA & display, predict & predict, decide & Secondary, AI-first \\ \hline
        \cite{Smith-Renner2020-hs} & model & NA & trust and reliance, system usability, decision satisfaction \& mental demand, understanding, feedback importance & predict, display & decide, modify & AI-first, User-guided adj. \\ \hline
        \cite{Riveiro2022-qi} & AI-system & efficacy & preferred type of explanation & predict & decide & AI-first \\ \hline
        \cite{Robbemond2022-ao} & ML model, AI system & efficacy, trust and reliance & trust and reliance, system usability, efficiency & predict, display & decide & AI-first \\ \hline
        \cite{Gaube2023-xe} & AI-based model (CHEST-AI) & efficacy & decision satisfaction \& mental demand, efficacy & predict, display & decide & AI-first \\ \hline
        \cite{Bhattacharya2023-im} & ML model & efficacy & system usability, trust and reliance, understanding, perceived explanation quality & predict, display & decide & AI-first \\ \hline
        \cite{Cabitza2023-ik} & AI & efficacy, trust and reliance & trust and reliance, perceived explanation quality & predict, display & predict, decide & AI-follow, AI-first \\ \hline
        \cite{Naiseh2023-ps} & AI tool & efficacy, trust and reliance & trust and reliance, understanding,  & predict, display & decide & AI-first \\ \hline
        \cite{Du2022-od} & clinical decision support system  & trust and reliance & preference & provide, display & predict, decide & Secondary \\ \hline
        \cite{Jiang2022-kt} & AI & user epistemic uncertainty (unclear responses) & system usability & request, predict, display, provide & collect, decide & AI-guided dial. \\ \hline
        \cite{Fogliato2022-uw} & AI & efficacy, efficiency, trust and reliance & decision satisfaction and mental demand, system usability & predict, display & predict, decide & AI-follow, AI-first \\ \hline
        \cite{Panigutti2022-va} & AI & trust and reliance & decision satisfaction \& mental demand, efficiency, system usability, trust and reliance, understanding, perceived explanation quality & provide, display & predict, decide & Secondary \\ \hline
        \cite{Calisto2022-bx} & assistant & efficacy, trust and reliance, efficiency & decision satisfaction \& mental demand, efficiency, fairness, system usability & predict, display & decide, request & AI-first, Request-driven \\ \hline
        \cite{Van_Berkel2022-uu} & AI & efficacy, trust and reliance, efficiency & system usability & predict, display & decide, provide & Others \\ \hline
        \cite{Suresh2022-xp} & model & trust and reliance & perception (open) & predict, display & decide, predict, modify & AI-first, Secondary, User-guided adj. \\ \hline
        \cite{Lam_Shin_Cheung2022-ho} & classification tool & efficacy, efficiency & NA & provide, display & request, predict & Secondary, Request-driven \\ \hline
        \cite{Hwang2022-dm} & AI model & efficacy & efficacy, trust and reliance & predict, display & decide & AI-first \\ \hline
        \cite{Matthiesen2021-ee} & prediction tool, algorithm & trust and reliance & efficiency, system usability & provide, display & predict, decide & Secondary \\ \hline
        \cite{Van_der_Waa2021-fz} & artificial agent & NA & fairness, quality and usefulness of explanations & predict, display, provide, delegate & decide, delegate, predict & AI-first, Delegate, Request-driven \\ \hline
        \cite{Lee2021-lu} & system & trust and reliance & decision satisfaction \& mental demand, system usability, trust and reliance & predict, display & modify, decide & AI-first, User-guided adj. \\ \hline
        \cite{Lindvall2021-lm} & Rapid Assisted Visual Search, AI & efficacy, efficiency & trust and reliance & predict, display & decide & Others \\ \hline
        \cite{Jacobs2021-ic} & system & efficacy & decision satisfaction \& mental demand, system usability & predict, display & decide & AI-first \\ \hline
        \cite{Tschandl2020-jv} & AI & trust and reliance, efficacy & NA &  predict, display & predict, decide & AI-follow \\ \hline
        \cite{Schaekermann2020-eo} & AI assistant & efficacy, efficiency & decision satisfaction \& mental demand, system usability, trust and reliance & predict, display & decide & AI-first \\ \hline
        \cite{Van_den_Brandt2020-fu} & predicted MD & trust and reliance & decision satisfaction \& mental demand, trust and reliance & provide, display & predict & Secondary \\ \hline
        \cite{Porat2019-ak} & decision aid  & NA & system usability & predict, display & predict, modify, decide & Others \\ \hline
        \cite{Wang2022-jm} & NA & efficacy, efficiency & decision satisfaction \& mental demand & predict, display & decide & AI-first \\ \hline
        \cite{Reverberi_C2022-cc} & AI & efficacy, trust and reliance & efficacy, trust and reliance & provide, predict & predict, decide & Others \\ \hline
        \cite{Jungmann_F2023-lk} & AI assistance & efficacy & trust and reliance & predict, display & predict, decide & AI-follow \\ \hline
        \cite{Gu2023-zl} & xPath & efficacy & trust and reliance, understanding & predict, provide, display & request, modify, decide & AI-first, Request-driven, User-guided adj. \\ \hline
        \cite{Schemmer2023-ka} & AI & efficacy & decision satisfaction \& mental demand & predict, display & predict, decide & AI-follow \\ \hline
        \cite{Brachman2022-dq} & AI & efficacy, efficiency, trust and reliance & trust and reliance & predict, display & decide & AI-first, Others \\ \hline
        \cite{Desmond2021-gw} & AI & efficacy, trust and reliance, efficiency & NA & predict, display & decide, request & AI-first, Request-driven \\ \hline
        \cite{Ashktorab2021-us} & AI & efficacy, efficiency, trust and reliance & system usability, decision satisfaction \& mental demand & provide & predict, modify & Secondary, User-guided adj. \\ \hline
        \cite{Schrills2020-km} & AI & NA & trust and reliance, understanding & predict, display & decide & AI-first \\ \hline
        \cite{Mackeprang2019-dz} & Recommender system & efficacy, trust and reliance, efficiency & system usability, decision satisfaction \& mental demand, fairness & provide, predict, display & request, decide & AI-first, Request-driven \\ \hline
        \cite{Bernard2018-tx} & Visual-interactive interface & efficacy, efficiency & system usability & display & predict & Secondary \\ \hline
        \cite{Zytek2022-hr} & Sibyl  & NA & trust and reliance, system usability & provide, display & modify, predict, request & Secondary, Request-driven, User-guided adj. \\ \hline
        \cite{Wang2021-ra} & machine learning  & objective understanding, trust and reliance & subjective understanding & predict, display & predict, decide & AI-follow \\ \hline
        \cite{Grgic-Hlaca2019-cu} & computer program & trust and reliance, efficacy & decision satisfaction \& mental demand & predict, display & predict, decide & AI-follow \\ \hline
        \cite{Kahr2023-yh} & AI system & trust and reliance & trust and reliance & predict, display & decide & AI-first\\ \hline
        \cite{Khadpe2020-vy} & bot & efficiency & system usability & request, predict & request, collect, decide & Request-driven, AI-guided dial. \\ \hline
        \cite{Leichtmann2023-xx} & Forestly & trust and reliance & understanding, trust and reliance & predict, display & decide, provide & AI-first \\ \hline
        \cite{Morrison2023-rr} & AI  & trust and reliance, efficacy & decision satisfaction \& mental demand, system usability & predict, display & predict, decide & AI-follow \\ \hline
        \cite{Tolmeijer2022-hm} & AI support & trust and reliance & fairness, trust and reliance & predict & predict, decide & AI-follow , AI-first\\ \hline
        \cite{Wu2022-zy} & virtual AI & trust and reliance & efficacy, trust and reliance & predict & predict, decide & Others \\ \hline
        \cite{Bucinca2021-cv} & AI & trust and reliance, efficacy & decision satisfaction \& mental demand, system usability, trust and reliance & predict, display & request, decide, predict & AI-follow, AI-first, Request-driven \\ \hline
        \cite{Fan2022-xt} & AI & efficacy & system usability, understanding, trust and reliance & predict, display & decide & Others \\ \hline
        \cite{Hofeditz2022-fq} & AI & NA & subjective comments & predict, display & decide & AI-first \\ \hline
        \cite{Peng2022-xt} & AI & efficacy, trust and reliance & NA & predict & decide & AI-first \\ \hline
        \cite{banas2022} & AI & NA & efficacy, trust and reliance & predict & decide & Others \\ \hline
        \cite{Molina2022-ms} & AI & NA & trust and reliance, understanding & predict, display & request, decide, modify & AI-first, Request-driven, User-guided adj. \\ \hline
        \cite{Rechkemmer2022-yy} & ML model & trust and reliance & efficacy, trust and reliance & predict, display & predict, decide & AI-follow \\ \hline
        \cite{Lai2022-ye} & model & efficacy & decision satisfaction \& mental demand, understanding & provide, display & predict, request, decide & Secondary, Request-driven \\ \hline
        \cite{Lu2021-xi} & ML model & trust and reliance & efficacy, trust and reliance, understanding & predict, display & predict, decide & AI-follow \\ \hline
        \cite{Bunde2021-of} & AI & NA & trust and reliance, system usability & predict, display & decide, provide & AI-first \\ \hline
        \cite{Nguyen2018-ao} & system, model & efficacy & NA & predict, provide, display & predict, decide, modify & AI-follow, User-guided adj. \\ \hline
        \cite{M_A_Rahman2021-pk} & NA & NA & trust and reliance & predict, provide & predict, decide & Others \\ \hline
        \cite{Appelganc2022-kv} & AI system, DSS, colleague & NA & trust and reliance & predict & decide & AI-first \\ \hline
        \cite{Cabrera2023-kf} & AI & accuracy & decision satisfaction \& mental demand, trust and reliance, system usability & predict, display & decide & AI-first \\ \hline
        \cite{Cau2023-nc} & AI & efficacy, trust and reliance & decision satisfaction \& mental demand, trust and reliance & predict, display & decide & AI-first \\ \hline
        \cite{Alufaisan2021-yt} & AI & trust and reliance, efficiency, efficacy & decision satisfaction \& mental demand, self-reported decision-making strategies & predict, display & decide & AI-first \\ \hline
        \cite{Liu2021-gm} & AI & efficacy, trust and reliance & system usability & predict, display & decide, request, modify & AI-first, Request-driven, User-guided adj. \\ \hline
        \cite{Bansal2021-xt} & Marvin & efficacy, trust and reliance & system usability & predict, display & decide & AI-first \\ \hline
\end{xltabular}

\bibliographystyle{ACM-Reference-Format}
\bibliography{sample-base}